\documentclass[twocolumn,aps,pra,longbibliography]{revtex4-2}
\usepackage[utf8]{inputenc}
\usepackage{url}
\usepackage{multirow}
\usepackage{amsmath}
\usepackage{amssymb}
\usepackage{graphicx}
\usepackage[justification=justified,size=small]{caption}
\usepackage{ragged2e}
\usepackage{subcaption}
\usepackage{mathrsfs} 
\usepackage{float}
\usepackage{bm}
\usepackage{physics}
\usepackage{mathtools}
\usepackage{array}
\usepackage{setspace}
\usepackage{ulem}
\usepackage[thicklines]{cancel}
\usepackage{newfloat,algcompatible}
\usepackage{etoolbox}
\usepackage{ragged2e}
\usepackage[utf8]{inputenc} 
\usepackage[T1]{fontenc}

\AtEndEnvironment{algorithm}{\noindent\hrulefill\par\nobreak\vskip-8pt}

\DeclareFloatingEnvironment[
    fileext=loa,
    listname=List of Algorithms,
    name=\textbf{Algorithm},
    placement=tbhp,
]{algorithm}
\DeclareCaptionFormat{algorithms}{\vskip-15pt\hrulefill\par#1#2#3\vskip-6pt\hrulefill}
\algblock[Input]{Input}{EndInput}
\algblockdefx[Input]{Input}{EndInput}%
    [1]{\textbf{Input} #1}%
    {}

\usepackage[unicode=true,bookmarks=true,bookmarksnumbered=false,bookmarksopen=false,
 breaklinks=false,pdfborder={0 0 1},backref=false,colorlinks=true]
 {hyperref}
\hypersetup{
 linkcolor=magenta, urlcolor=blue, citecolor=blue, pdfstartview={FitH}, unicode=true}
\newtheorem{theorem}{Theorem}
\newtheorem{proof}{Proof}

\newcommand{\sust}{Department of Physics, Southern University of Science and Technology, Shenzhen, 518055, China}

\newcommand{\cdut}{Department of Physics, College of Physics, Chengdu University of Technology, Chengdu, 610059, China}
\begin{document}

\title{Channel Superposition Mitigates Photon Loss Errors in Quantum Illumination}
\author{Fei Li}

\email{12031329@mail.sustech.edu.cn}
\affiliation{\sust}

\author{Xiao-Wei Li}
\email{lixiaowei@cdut.edu.cn}
\affiliation{\cdut}

\author{Oscar Dahlsten}
\email{oscar.dahlsten@cityu.edu.hk}
\affiliation{Department of Physics, City University of Hong Kong, Hong Kong SAR, 518057, China}
\affiliation{Institute of Nanoscience and Applications, Southern University of Science and Technology, Shenzhen, 518055, China}

\date{\today}
\begin{abstract}
In quantum illumination, the probe photon is entangled with an ancilla photon, and both are jointly measured at the end. 
The entanglement between the probe and ancilla photons enhances the detection performance per unit average photon number in the probe mode, particularly in low-reflectivity and high-noise scenarios.
However, photon loss severely limits the practical advantage of such protocols. To address this, we employ a channel superposition framework, which encompasses two kinds of channel superposition protocols: indefinite causal order (ICO) and path superposition with disjoint environment (PS-DE).  
Our analytical and numerical analysis based on the quantum Chernoff bound shows that both ICO and PS-DE can, in principle, achieve an advantage. The advantage persists as long as non-zero interference remains, reverting to the performance of standard quantum illumination once the interference is completely suppressed. 
Crucially, the ICO protocol is significantly more robust, maintaining a tighter upper bound on the error probability than standard quantum illumination and the PS-DE approach. 
This performance hierarchy is rooted in their fundamental structures: ICO exploits a shared environment to generate stronger quantum interference, while PS-DE, relying on disjoint environments, offers a more experimentally tractable albeit less potent alternative.
\end{abstract}
\maketitle
\section{Introduction}
\begin{figure*}[!t]
    \centering
    \begin{subfigure}[b]{0.8\textwidth}
        \centering
        \includegraphics[width=\linewidth]{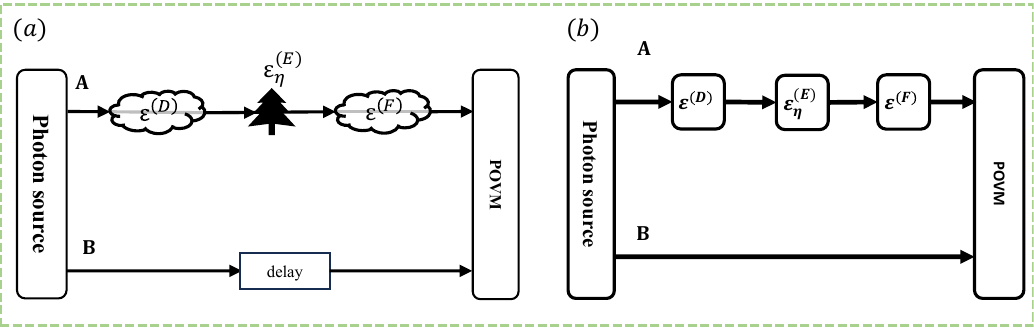} 
    \end{subfigure}

    \vspace{5pt} 
    \begin{subfigure}[b]{0.8\textwidth}
        \centering
        \includegraphics[width=\linewidth]{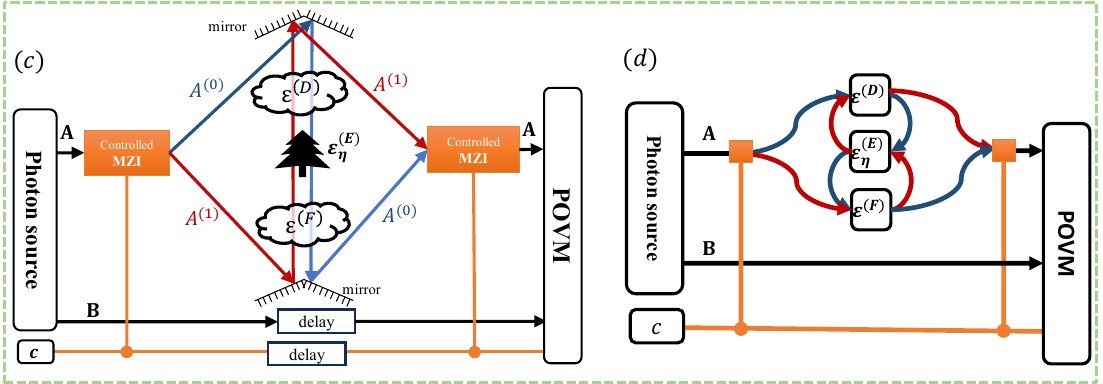}
    \end{subfigure}

    \vspace{5pt} 
    \begin{subfigure}[b]{0.8\textwidth}
        \centering
        \includegraphics[width=\linewidth]{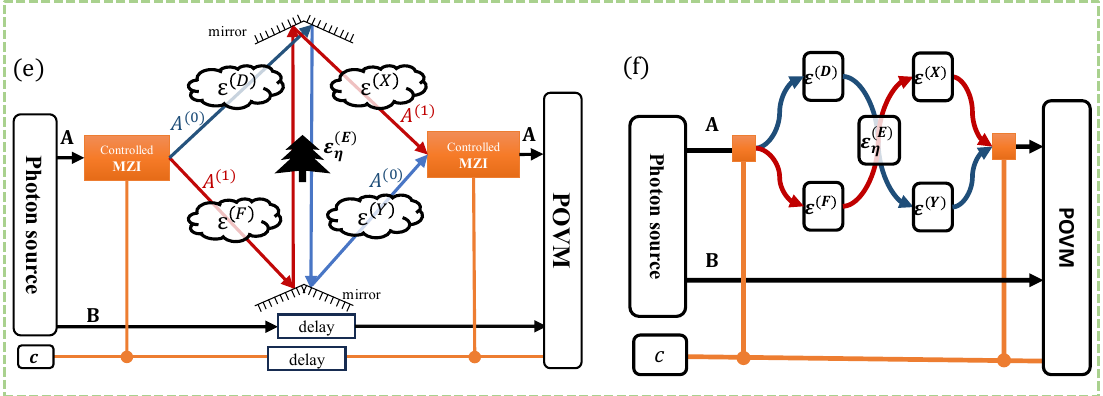}
    \end{subfigure}
    
    \caption{\textbf{Quantum illumination protocols under photon loss: QI, ICO, and PS-DE.} 
    \justifying{\textit{Key distinction:} Both ICO and PS-DE introduce quantum interference in the output state to enhance noise resilience relative to standard QI.
    ICO superposes causal orders through a shared environment, while PS-DE superposes spatial paths via disjoint, independent environments.
    (\textbf{a}\textendash\textbf{b}) \textit{Standard QI protocol.}
    (\textbf{a}): mode $A$ interacts with the target $\mathcal{E}^{(E)}_\eta$, 
    where it undergoes probabilistic photon loss, represented by $\mathcal{E}^{(D)}$ before and $\mathcal{E}^{(F)}$ after the detection event.
    Mode $B$ is retained as an ancilla for subsequent joint measurements. 
    (\textbf{b}): A schematic of the sequential process. Mode $A$ passes through  $\mathcal{E}^{(D)} \to\mathcal{E}^{(E)}_{\eta}\to\mathcal{E}^{(F)}$, followed by a joint measurement (POVM) together with mode $B$. 
    (\textbf{c}\textendash\textbf{d}) \textit{ICO protocol.}
    (\textbf{c}): Mode $A$ propagates through one of two paths selected by a Mach-Zehnder interferometer (MZI), with the control system modulating the phase shift by the state of the control qubit $\ket{\psi_c}$. If the state of the control qubit is $\ket{\psi_c}=\ket{0}$ (corresponding to path $A^{(0)}$), or $\ket{\psi_c}=\ket{1}$ (corresponding to path $A^{(1)}$), mode $A$ experiences photon losses in reverse orders, thereby enabling exploration of causal indefiniteness.
    Crucially, the two paths undergo coherent recombination at the end of the MZI. 
    (\textbf{d}) Schematic of the ICO protocol: The sequence of channel interactions for mode $A$ ($\mathcal{E}^{(D)}$, $\mathcal{E}^{(E)}_\eta$ and $\mathcal{E}^{(F)}$) is dictated by the state of the control qubit $\ket{\psi_c}$. 
    (\textbf{e}\textendash\textbf{f}) \textit{PS-DE protocol.} 
    (\textbf{e}): Mode $A$ undergoes path superposition via a controlled MZI, where each path experiences disjoint channel combinations (e.g., $A^{(0)}$ encounters $\mathcal{E}^{(D)}$ and $\mathcal{E}^{(Y)}$; $A^{(1)}$ encounters $\mathcal{E}^{(F)}$ and $\mathcal{E}^{(X)}$) before coherent recombination.  
    (\textbf{f}) Schematic of the PS-DE protocol: the channel processing for mode $A$ leverages quantum interference from superposing distinct paths/channels.
    } 
    }
    \label{fig:main}
\end{figure*}

Quantum illumination (QI)\cite{Lloyd_qillum2008} leverages quantum entanglement to detect low-reflectivity targets in environments with strong background noise~\cite{Lloyd_qillum2008,Yung2020Oneshot,PhysRevLett.114.080503,PhysRevA.89.062309,Ragy:14}. As depicted in Fig.~\ref{fig:main}(a), the probe system (mode A) is entangled with another system (mode B), and the final measurement involves both of those systems. QI has garnered significant theoretical and experimental attention due to its superiority over classical methods in high loss and strong thermal noise scenarios, where the exponential decay of error probability is governed by the Quantum Chernoff Bound (QCB)~\cite{Shapiro_2009,PhysRevLett.101.253601,PhysRevA.80.052310,PhysRevA.80.022320,PhysRevLett.110.153603,zhang_quantum_2014}. 
Despite advances in QI, photon loss remains a critical challenge in practical quantum systems, posing an obstacle to QI’s performance. Unlike classical protocols, which rely on classical signal amplification to mitigate loss, QI depends on entanglement between the probe and ancilla photons to retain non-classical correlations. However, photon loss disrupts this entanglement, degrading the quantum advantage in the QCB error exponent that quantifies the exponential decay of error probability~\cite{zhang_quantum_2014,PhysRevA.109.062440,noh2021quantumilluminationnongaussianstates}.
Developing innovative strategies to mitigate photon loss is imperative to unlock QI’s full potential.

Recently, the coherent control of quantum channels---where a quantum system traverses multiple noisy processes in a superposition controlled by an ancillary quantum system---has emerged as a strategy to mitigate noise and enhance information processing in quantum technologies~\cite{PhysRevLett.120.120502,8966996,Abbott2020communication,PhysRevA.107.062208,PhysRevLett.129.100603,PhysRevA.99.062317}. 
Among such strategies, we focus on two specific realizations of channel superposition: indefinite causal order (ICO) and a path superposition scheme with disjoint environments (PS-DE).
In ICO, the temporal order of quantum operations is coherently superposed, inducing quantum interference between distinct causal structures~\cite{PhysRevA.88.022318, Giulia_science2017,PhysRevLett.121.090503, Oreshkov2012Quantum,PhysRevA.88.022318,Procopio2015Experimental,PhysRevLett.113.250402,PhysRevResearch.2.033292,PhysRevLett.124.190503,PhysRevLett.129.230604}. A defining feature of ICO is its reliance on shared environmental modes, which couple to both operational sequences.
In contrast, in our PS-DE scheme, photons traverse multiple spatial paths in superposition, with each path subjected to an independent noise process. 
Crucially, PS-DE employs disjoint environmental modes, unlike the shared environment in ICO, ensuring that noise processes across different paths remain uncorrelated.
By exploiting interference between superpositioned channels, this approach can suppress effective noise and enhance signal distinguishability, offering advantages for noisy quantum metrology~\cite{kurdzialek_using_2023, chapeau-blondeau_indefinite_2022, chapeau-blondeau_noisy_2021} and quantum communication~\cite{ frey_indefinite_2019, goldberg_evading_2023}.
Despite these advances, the potential of ICO/PS-DE to mitigate photon loss errors in quantum illumination remains unexplored.

Our work addresses this gap through a systematic comparison of ICO and PS-DE within the QI framework. 
This comparative approach is motivated by a fundamental trade-off: while ICO is viable in controlled laboratory settings and achieves superior performance by leveraging shared environments, PS-DE offers a more readily scalable architecture for practical implementation, relying on path superposition with independent environments.
To this end, we introduce a control qubit, $c$, that generalizes the standard QI protocol by coherently controlling the order in which the probe photon interacts with environmental subsystems, specifically, $D$, $E$, and $F$, via both ICO and PS-DE configurations (Fig.~\ref{fig:main}).
Our analysis shows that both PS-DE ($D-E-Y$ vs $F-E-X$) and ICO ($D-E-F$ vs $F-E-D$) introduce quantum interference effects, which can enhance the performance of the QI protocol. Using the QCB, we derive analytical upper bounds on error probabilities, demonstrating that ICO-based protocols can achieve a more significant and robust advantage, outperforming both PS-DE and standard QI in highly lossy scenarios and, crucially, in the large-dimensional limit. The advantage persists even when the control qubit flip weakens the interference strength, provided that non-zero interference remains. Channel superposition can thus be viewed as a resource for achieving robust quantum illumination in lossy environments.

The remainder of this paper is structured as follows:
Section \ref{sec:unified_framework} introduces a unified framework for analyzing QI protocols, presenting the physical model and output states for standard QI, PS-DE, and ICO under a common mathematical structure. Subsection \ref{sec:quantum_illumination} details the model setup for standard QI with photon loss, including the initial state and the two-way probabilistic loss channel affecting the probe mode $A$. 
In Subsection \ref{sec:ICO}, we introduce the concept of channel superposition for QI. Then we give analytical expressions for the error probability in subsection~\ref{sec:QCB}.  
Section \ref{sec:resource} analyzes the resource theory of QI with channel superposition, examining the impact of control qubit noise (phase flip and bit flip) on the coherence that enables channel superposition. The effectiveness of channel superposition is quantified using the decoherence coefficient $\gamma$.
Simulation results are presented in Section \ref{sec:simulation}, which is organized into three subsections: Subsection \ref{sec:setup} details the numerical setup, including the initial state preparation and the choice of truncation dimension; Subsection \ref{subsec:comparison} provides a qualitative and quantitative comparison of the interference strength and resulting performance potential between ICO and PS-DE protocols; 
and subsection \ref{subsec:advantage} presents a comprehensive numerical analysis of the performance advantage and robustness of the ICO scheme across various loss and reflectivity regimes.
The simulations demonstrate the superior performance of ICO-enhanced systems in detection efficiency and reliability.
Section~\ref{sec:origin} compares the gain and cost of ICO with a third scheme wherein the environment modes are coherently controlled.
Finally, Section \ref{sec:conclusion} summarizes our findings and discusses future research directions.

\section{Quantum Illumination with ICO and PS-DE Protocols}
\label{sec:unified_framework}

This section establishes a unified framework to analyze and compare quantum illumination protocols. We begin by reviewing the standard QI protocol with photon loss in Section~\ref{sec:quantum_illumination}. We then generalize this setup in Section~\ref{sec:ICO} to incorporate channel superposition---comprising both PS-DE and ICO---and derive the general structure of the output state. Finally, in Section~\ref{sec:QCB}, we analyze the quantum Chernoff bound within this unified framework, providing the foundation for comparing the performance of the different protocols.

\subsection{Standard Quantum Illumination}\label{sec:quantum_illumination}

The goal of quantum illumination is to detect the presence of a low-reflectivity target in a high-noise environment. 
We follow the physical model for QI constructed by Ref.~\cite{zhang_quantum_2014}.
In that model, the target to be probed has a reflectivity $\eta$. Consider a bipartite quantum system defined on the Hilbert space $\mathcal{H}_A \otimes \mathcal{H}_B$, where $\mathcal{H}_A$ and $\mathcal{H}_B$ are the Hilbert spaces of the probe mode $A$ and the ancilla mode $B$, respectively. The initial state is prepared as an entangled state:
\begin{equation}
\ket{\psi}_{AB} = \sum_n c_n \ket{n}_A \otimes \ket{n}_B,
\end{equation}
where the coefficients $c_n$ are determined by the specific state preparation method.
It is assumed that mode $A$ experiences probabilistic photon loss before and after interacting with the target. The probabilistic loss channel $\mathcal{E}^{(a)}$ acting on $\rho_{AB}\in\mathcal{H}_A\otimes\mathcal{H}_B$ with survival probability $p^{(a)}$ can be expressed as~\cite{PhysRevA.78.032315,zhang_quantum_2014}:
\begin{equation}\label{eq:loss}
\mathcal{E}^{(a)}(\rho_{AB}) = p^{(a)}\rho_{AB} + \left(1-p^{(a)}\right)\ket{0}_A\bra{0} \otimes \rho_B.
\end{equation}
Here, the superscript $(a)$ explicitly labels the environmental system involved in the Stinespring dilation of this channel. Eq.~(\ref{eq:loss}) signifies that upon photon loss, the probe mode $A$ is replaced by the vacuum state $\ket{0}_A$, while the reduced state of the ancilla mode $B$ is updated to $\mathrm{Tr}_A(\rho_{AB})$, reflecting the decoherence induced by the loss.

We denote the probe process channel as $\mathcal{E}_{\eta}^{(E)}$, characterized by the target's reflectivity $\eta$. When the target is absent, $\eta=0$. When the target is present, mode $A$ is reflected with a relatively low probability $\eta$.
The traditional quantum illumination protocol (Fig.~\ref{fig:main} (a)-(b)) provides the foundational framework for our analysis. In this protocol, mode $A$ interacts with the target via a beam splitter of reflectivity $\eta$ that couples it with an environmental mode $E$. The unitary operation $V_{AE}(\eta)$ describing this beam splitter interaction is~\cite{zhang_quantum_2014}:
\begin{equation}
    V_{AE}(\eta):=\exp\left\{
    \arctan\left[\sqrt{(1-\eta)/\eta}\right](a_A a_E^\dagger-a_A^\dagger a_E )\right\},
\end{equation}
where $a_A^\dagger$ and $a_E^\dagger$ denote the creation operators for modes $A$ and $E$, respectively. The environment is in a thermal state $\rho_E$ with mean photon number $N$:
\begin{equation}\label{eq:th}
    \rho_E=\sum_{r=0}^\infty \frac{N^r}{(N+1)^{r+1}} \ket{r}_E\bra{r}.
\end{equation}
The resulting channel $\mathcal{E}_\eta^{(E)}$ acting on the system $\rho_{AB}$ is given by:
\begin{equation}
\begin{aligned}
\mathcal{E}_{\eta}^{(E)}(\rho_{AB}) &= \mathrm{Tr}_E\left[ V_{AE}(\eta) (\rho_{AB} \otimes \rho_E) V_{AE}(\eta)^\dagger \right]\\
&= \eta\rho_{AB}+(1-\eta)\rho_A^{\text{th}}\otimes\rho_B,
\end{aligned}
\end{equation}
where $\rho_A^{\text{th}}$ is a thermal state in mode $A$ with mean photon number $N$, described by the density matrix:
\begin{equation}
    \rho_A^{\text{th}} = \sum_{r=0}^\infty \frac{N^r}{(N+1)^{r+1}} \ket{r}_A\bra{r}.
\end{equation}

Combining $\mathcal{E}_{\eta}$ with two-way probabilistic losses, the output state can be written as:
\begin{equation}\label{eq:qi_out}
    \rho_{\mathrm{\text{out}}}^{\text{QI}}(\eta)=\mathcal{E}^{(F)}\circ\mathcal{E}^{(E)}_{\eta}\circ\mathcal{E}^{(D)}(\rho_{AB}),
\end{equation}
where $\rho_{AB}=\ket{\psi}_{AB}\bra{\psi}$, $\mathcal{E}^{(D)}$ and $\mathcal{E}^{(F)}$ are the loss channels before and after the probe, respectively.
Subsequently, we perform a joint measurement on mode $A$ and mode $B$ to discriminate whether the target exists, i.e., to determine whether $\eta=0$ or $\eta>0$.

The minimum error probability for distinguishing between the two hypotheses (target present or absent) over $M$ repeated and independent detections in a QI protocol is denoted by $P_{\text{err},M}$. It is given by the Helstrom formula for multiple copies~\cite{zhang_quantum_2014}: 
\begin{equation}\label{eq:P_err} P_{\text{err},M}=\frac{1}{2}\left\{1-\frac{1}{2}\left\|\rho_{\mathrm{\text{out}}}(\eta)^{\otimes M}-\rho_{\mathrm{\text{out}}}(0)^{\otimes M}\right\|\right\},
\end{equation}
where $\rho_{\mathrm{\text{out}}}(\eta)$ denotes the output single-copy state when the target is present (with reflectivity $\eta$), and
$\rho_{\mathrm{\text{out}}}(0)$ indicates the output state when the target is absent. Here $||\cdot||$ is the trace norm, defined as $||A||=\frac{1}{2}\tr(|A|)$, with $|A|=\sqrt{A^\dagger A}$.
For the standard QI protocol,
$P_{\text{err}, M}^{\text{QI}}$ is bounded above by the quantum Chernoff bound (QCB)~\cite{PhysRevLett.98.160501,10.1214/08-AOS593}, which provides the asymptotic limit of the error probability for large $M$:
\begin{equation}\label{eq:QCB}
    P_{\mathrm{\text{err}},M}^{\text{QI}}\leq P_{\text{QCB},M}^{\text{QI}} = e^{-\frac{1}{2}M\epsilon^{\text{QI}}},
\end{equation}
with 
\begin{equation}
    \epsilon^{\text{QI}} = -\ln \left\{Q\left[\rho_{\text{out}}^{\text{QI}}(\eta), \rho_{\text{out}}^{\text{QI}}(0)\right]\right\}.
\end{equation}
Here, we adopt the definition from Ref.~\cite{PhysRevLett.98.160501}, where the \textit{Q function} for any two quantum states $\rho$ and $\rho'$ is given by:
\begin{equation}
Q(\rho, \rho') := \min_{0 \leq s \leq 1} \mathrm{Tr}\left[\rho^s \rho'^{1-s}\right].
\label{eq:Q_function_definition}
\end{equation}
This function quantifies the operational distinguishability between $\rho$ and $\rho'$ according to \cite{PhysRevLett.98.160501}.

Notice that QCB evaluation only involves single-copy states $\rho_{\text{out}}(\eta)$ and $\rho_{\text{out}}(0)$, simplifying the computational complexity in comparison to the approach described in Eq.~\eqref{eq:P_err}.

\subsection{Channel Superposition Protocols: ICO and PS-DE}\label{sec:ICO}

\begin{figure*}[t]
    \centering
    \subfloat[Circuit of ICO]{
        \begin{minipage}[t]{0.48\textwidth} 
            \centering                    \includegraphics[width=\linewidth]{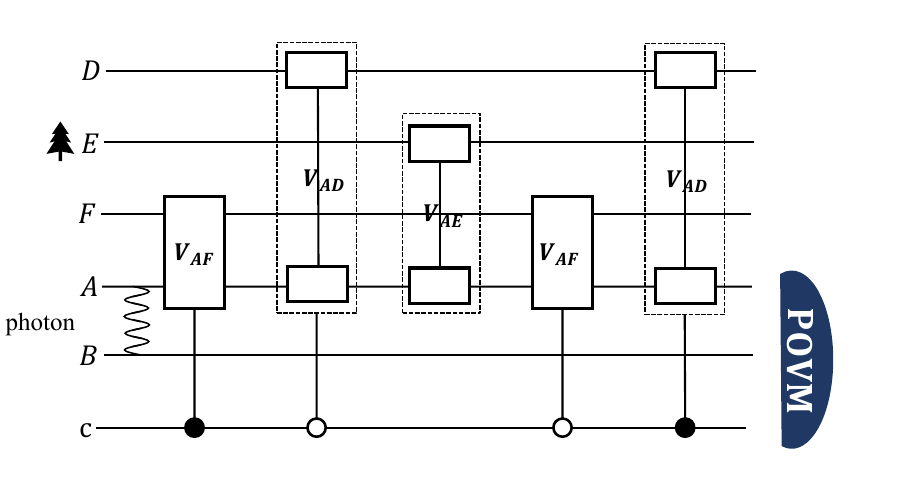} 
        \end{minipage}
        \label{fig:subfig1}
    }
    \subfloat[Circuit of PS-DE]{
        \begin{minipage}[t]{0.48\textwidth}  
            \centering
            \includegraphics[width=\linewidth]{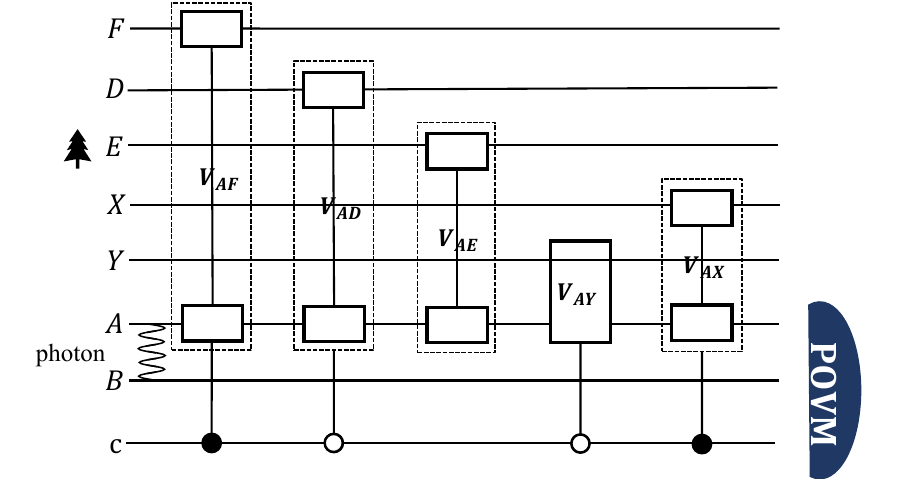}
        \end{minipage}
        \label{fig:subfig2}
    }
    \caption{\textbf{Schematic diagram of a unitary circuit.}
    \justifying{(a)\textit{ICO protocol}. Photon A interacts with three auxiliary modes (D, E, F) via unitary operators $V_{AD}$, $V_{AE}$, $V_{AF}$, with causal order controlled by qubit $c$.
    When the control qubit $c=\ket{0}$, the photon travels through paths D, E, and F in sequence (i.e., following the order D $\to$ E $\to$ F); when $c=\ket{1}$, the photon travels through paths F, E, and D in reverse order (i.e., following the order F $\to$ E $\to$ D).
    (b)\textit{PS-DE protocol}. Photon A interacts with five ancilla modes (D, F, E, X, Y) through unitary operators $V_{AD}$, $V_{AF}$, $V_{AE}$, $V_{AX}$, and $V_{AY}$. When the control qubit $c=\ket{0}$, the photon traverses paths D, E, and Y; when $c=\ket{1}$, the photon traverses paths F, E, and X.}}
    \label{fig:circuit}
\end{figure*}
We explore the applications of two specific channel superposition protocols, ICO and PS-DE, in quantum illumination, focusing on their robustness against photon loss. Both schemes are implemented by introducing an additional control qubit $\ket{\psi}_c$ to control the path of mode $A$ coherently.

Specifically, for the ICO protocol, if $\ket{\psi}_c=\ket{0}$, mode $A^{(0)}$ traverses through channel $\mathcal{E}^{(D)}$ before target detection $\mathcal{E}^{(E)}_\eta$ and then through $\mathcal{E}^{(F)}$; conversely, if $\ket{\psi}_c=\ket{1}$, mode $A^{(1)}$ first passes through $\mathcal{E}^{(F)}$, target $\mathcal{E}^{(E)}_\eta$, and subsequently through $\mathcal{E}^{(D)}$.
The schematic diagram and circuit of the proposal are presented in Fig.~\ref{fig:main} (c)-(d), respectively. Herein, the control qubit is coupled to the optical paths to coherently control the order of channel operations, realizing the ICO protocol. The potential experimental configuration for our proposal is detailed in Appendix~\ref{app:experimental_setup}. 

For the PS-DE protocol, when $\ket{\psi}_c = \ket{0}$, mode $A^{(0)}$ traverses through loss channel $\mathcal{E}^{(D)}$ before target detection $\mathcal{E}^{(E)}_\eta$ and then through loss channel $\mathcal{E}^{(Y)}$; when $\ket{\psi}_c = \ket{1}$, mode $A^{(1)}$ traverses through loss channel $\mathcal{E}^{(F)}$, target channel $\mathcal{E}^{(E)}_\eta$, and subsequently through $\mathcal{E}^{(X)}$.
The schematic diagram and circuit of the proposal are presented in Fig.~\ref{fig:main} (e)-(f).

The channels can be represented using Kraus operators as follows:
\begin{equation}
\begin{aligned}
        \mathcal{E}^{(a)}(\rho)&=\sum_{i}K_i^{(a)}\rho K_i^{(a) \dagger},~a\in\{D,F, X, Y\},\\
        \mathcal{E}_{\eta}(\rho)&=\sum_{ij}K_{ij}^{(E)}\rho K_{ij}^{(E) \dagger},
\end{aligned}
\end{equation}
where $\sum_i K_i^{(a)\dagger} K_i^{(a)}= \mathbb{I}_{A}$ and $\sum_{ij} K_{ij}^{(E)\dagger} K_{ij}^{(E)} =\mathbb{I}_{A}$ ensure complete positivity and trace preservation.
The lossy channel $\mathcal{E}^{(a)}$ with survival probability $p^{(a)}$ admits a Kraus representation with operators:
\begin{equation}
\begin{split}
    K^{(a)}_0 & = \sqrt{p^{(a)}} \mathbb{I}_{A}, \\
    K^{(a)}_n & = \sqrt{1-p^{(a)}}\ket{0}_A\bra{n-1}, \quad \text{for } 1\leq n\leq \mathcal{D},
\end{split}
\end{equation}
where $\mathcal{D}$ is the Hilbert space truncation dimension, numerically determined in Section~\ref{sec:setup} to ensure convergence.

Formally, both protocols can be rigorously defined within the Stinespring dilation framework, where the channels are implemented by coupling the system to ancillary environmental modes via global unitary operations, followed by partial trace.
The initial state of the entire system (control, probe, ancilla, and all environmental modes) is given by:
\begin{equation}
\rho_\text{init}^\text{u}=\rho_{cAB}\otimes\ket{0}_\text{auxiliary(u)}\bra{0}\otimes\rho_E,
\end{equation}
where $\text{auxiliary(ICO)}=\{D,F\}$ and $\text{auxiliary(PS-DE)}=\{D,F,X,Y\}$.
The key difference between ICO and PS-DE lies in their global unitary operations $V^{\text{u}}$.
For the ICO Protocol, the unitary $V^{\text{ICO}}$ entangles the control qubit with the \textit{same} set of environmental modes ($D, E, F$), as expressed by
    \begin{equation}
    \begin{aligned}
        V^\text{ICO}&=\ket{0}_c\bra{0}\otimes V_{AF}V_{AE}V_{AD}\otimes\mathbb{I}_B\\
        &+\ket{1}_c\bra{1}\otimes V_{AD}V_{AE}V_{AF}\otimes\mathbb{I}_B.
    \end{aligned}
    \end{equation}
For the PS-DE protocol, we choose a unitary
    \begin{equation}
    \begin{aligned}
        V^\text{PS-DE}&=\ket{0}_c\bra{0}\otimes V_{AY}V_{AE}V_{AD}\otimes\mathbb{I}_{BFX}\\
        &+\ket{1}_c\bra{1}\otimes V_{AX}V_{AE}V_{AF}\otimes\mathbb{I}_{BDY},
    \end{aligned}
    \end{equation}
that entangles the control qubit with \textit{disjoint} sets of environmental modes $(D,E,F,X,Y)$.
This choice of disjoint environments for PS-DE reflects independent noise processes in each path.
A corresponding unitary circuit representation of this process is depicted in Fig.~\ref{fig:circuit}.
The final system-only output state $\rho_{\text{out}}^{\text{u}}$ is then obtained by tracing out all environmental modes ($D, E, F, X, Y$) after the global unitary evolution:
\begin{equation}
\rho_{\text{out}}^{\text{u}} = \Tr_{\text{auxiliary(u)},E} \left[ V^{\text{u}} \, \rho_{\text{init}}^\text{u} \, V^{\text{u}^\dagger} \right].
\end{equation}

From this Stinespring dilation framework, the effective channel on the system (control, probe, and ancilla) is obtained by performing the partial trace over all environmental modes. This process yields the Kraus operator representations for the overall channels of the ICO and PS-DE protocols:
\begin{equation}\label{eq:channel_kraus}
    \begin{aligned}
        W^{\text{ICO}}_{ijkl} & = \ket{0}_c\bra{0}\otimes K_i^{(F)}K_{jk}^{(E)} K_l^{(D)}\otimes \mathbb{I}_{B} \\
        & \quad + \ket{1}_c\bra{1}\otimes K_l^{(D)}K_{jk}^{(E)} K_i^{(F)}\otimes \mathbb{I}_B, \\
        W^{\text{PS-DE}}_{ii'jkll'} & = \left( \delta_{i'0}\delta_{l'0}\ket{0}_c\bra{0}\otimes K_i^{(Y)}K_{jk}^{(E)} K_l^{(D)}\otimes \mathbb{I}_{B} \right. \\
        & \left. + \delta_{i0}\delta_{l0}\ket{1}_c\bra{1}\otimes K_{i'}^{(X)}K_{jk}^{(E)} K_{l'}^{(F)}\otimes \mathbb{I}_B \right),
    \end{aligned}
\end{equation}
The detailed derivation from the global unitaries $V^{\text{u}}_{\text{global}}$ to these Kraus operators is provided in Appendix~\ref{app:Kraus}. For simplicity, and to focus on the effect of superposition itself, we assume identical loss channels: $p^{(D)}=p^{(F)}=p^{(X)}=p^{(Y)}=p$, implying $K^{(D)}_i=K_i^{(F)}=K_i^{(X)}=K_i^{(Y)} \equiv K_i$.

The output states for the two protocols are obtained by applying their respective quantum channels to the initial product state $\rho_c \otimes \rho_{AB}$, where the control qubit is initialized in $\rho_c = \ket{+}_c\bra{+}$ with $\ket{\pm}_c := (\ket{0}_c \pm \ket{1}_c)/\sqrt{2}$:
\begin{equation}
    \begin{aligned}
        \rho_{\text{out}}^{\text{ICO}} & = \sum_{ijkl} W_{ijkl}^{\text{ICO}} (\rho_c \otimes \rho_{AB}) W_{ijkl}^{\text{ICO}\dagger}, \\
        \rho_{\text{out}}^{\text{PS-DE}}  & = \sum_{ii'jkll'} W_{ii'jkll'}^{\text{PS-DE}} (\rho_c \otimes \rho_{AB}) W_{ii'jkll'}^{\text{PS-DE}\dagger}.
    \end{aligned}
\end{equation}
A key observation from our comparative analysis is that, despite their different Kraus representations, the output states for both protocols can be expressed in an identical mathematical form within the control qubit basis $\{ \ket{0}_c, \ket{1}_c \}$:
\begin{equation}
\rho_{\text{out}}^{\text{u}} (\eta) = \frac{1}{2} \begin{bmatrix}
\rho_{\text{out}}^{\text{QI}} (\eta) & \sigma^{\text{u}}(\eta) \\
\sigma^{\text{u}}(\eta) & \rho_{\text{out}}^{\text{QI}}(\eta)
\end{bmatrix}_c,
\label{eq:unified_output_structure}
\end{equation}
where $\text{u} \in \{ \text{ICO}, \text{PS-DE} \}$, $\rho_{\text{out}}^{\text{QI}}(\eta) = \mathcal{E}^{(F)} \circ \mathcal{E}_\eta^{(E)} \circ \mathcal{E}^{(D)}(\rho_{AB})$ is the output state of the standard quantum illumination protocol, and the off-diagonal term $\sigma^{\text{u}}(\eta)$ encodes the quantum interference effects characteristic of each specific superposition scheme.

The protocols differ solely in the specific construction of the interference term $\sigma^{\text{u}}$:
\begin{itemize}
\item \textbf{ICO}: The interference term results from a symmetric sum over Kraus operators:
\begin{equation}\label{eq:sigma_ICO}
\sigma^{\text{ICO}}(\eta) = \sum_{i,j} K_i \mathcal{E}_\eta^{(E)}\left(K_j \rho_{AB} K_i^\dagger\right) K_j^\dagger,
\end{equation}
where $\{K_i\}$ are the Kraus operators for the identical loss channels $\mathcal{E}^{(D)}$ and $\mathcal{E}^{(F)}$.
\item \textbf{PS-DE}: The interference term allows for a more general, asymmetric summation:
\begin{equation}\label{eq:sigma_ps}
\begin{aligned}
\sigma^{\text{PS-DE}}(\eta) &=\sum_{i,i',l,l'}\delta_{i0}\delta_{l0}\delta_{i'0}\delta_{l'0}K_i \mathcal{E}_\eta^{(E)}\left(K_l \rho_{AB} K_{l'}^\dagger\right) K_{i'}^\dagger\\
&=K_0\mathcal{E}_\eta^{(E)}\left(K_0\rho_{AB}K_0^\dagger\right)K_0^\dagger\\
&=p^2\mathcal{E}_\eta^{(E)}(\rho_{AB}).
\end{aligned}
\end{equation}
\end{itemize}


This architectural difference—shared versus separated environments—suggests that the two protocols may respond differently to photon loss. The following sections will explore the analytical and numerical implications of this fundamental distinction.

\subsection{Performance Analysis via Quantum Chernoff Bound}\label{sec:QCB}
\begin{figure}[t]
    \centering
    \includegraphics[width=\linewidth]{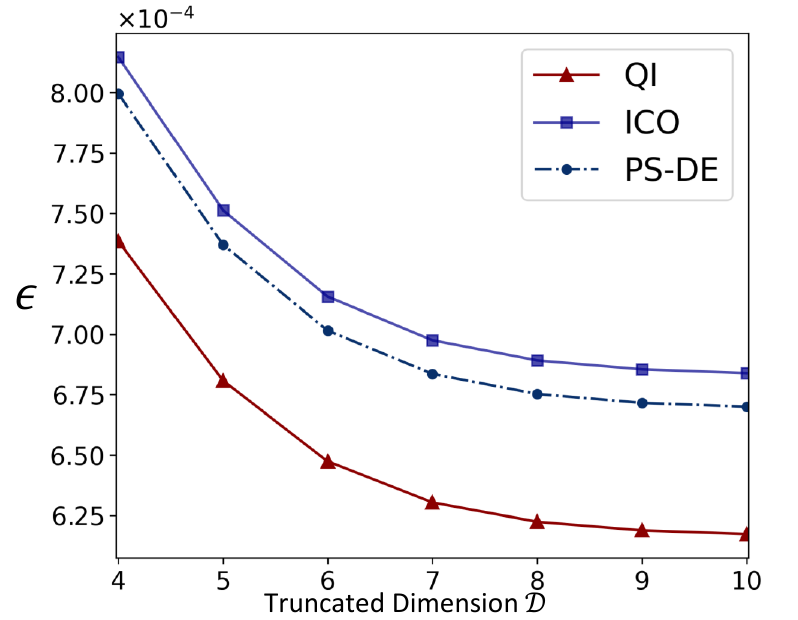}    
    \caption{\textbf{Impact of Truncated Dimension on $\epsilon$ for QI, PS-DE, and ICO Protocols.}
    \justifying{This figure demonstrates the relationship between the truncated dimension  $\mathcal{D}$ and the performance metrics $\epsilon^{\text{QI}}$, $\epsilon^{\text{PS-DE}}$, and $\epsilon^{\text{ICO}}$. The truncated dimension represents the maximum number of states considered in a quantum system, serving as an approximation technique to manage computational complexity in quantum simulations. 
    As $\mathcal{D}$ increases, $\epsilon^{\text{QI}}$, $\epsilon^{\text{PS-DE}}$, and $\epsilon^{\text{ICO}}$ converge, highlighting the diminishing impact of truncation on simulation accuracy at higher dimensions.
    The simulation parameters include a target reflectivity $\eta = 0.1$, an average photon number $N=0.5$ for mode $E$, an invariant loss channel probability $p = 0.8$, and an average photon number $N_t=0.01$ for the transmitted mode $A$.}}
    \label{fig:dimension}
\end{figure}
Due to their shared mathematical structure, the performance analysis of both ICO and PS-DE protocols follows a similar approach, ultimately reducing to the same type of optimization problem over the parameter $s$.
To evaluate this, we apply a Hadamard transformation to the control qubit such that it is in the basis ($\{\ket{+}_c, \ket{-}_c\}$), which block-diagonalizes the output state:
\begin{equation}\label{eq:block_diagonal_form}
\rho_{\text{out}, \pm}^{\text{u}}(\eta) = \frac{1}{2} \begin{bmatrix}
\rho_{\text{out}}^{\text{QI}}(\eta) + \sigma^{\text{u}}(\eta) & 0 \\
0 & \rho_{\text{out}}^{\text{QI}} - \sigma^{\text{u}}(\eta)
\end{bmatrix}_c.
\end{equation}
The QCB for a protocol yielding the generalized state $\rho_{\text{out}}^{\text{u}}$ is following Eq.~(\ref{eq:QCB}):
\begin{equation}
P_{\text{err}, M}^{\text{u}} \leq P_{\text{QCB},M}^{\text{u}}=\exp\left(-\frac{1}{2} M \epsilon^{\text{u}}\right), 
\end{equation}
where 
\begin{equation}\label{eq:trace_cqubit}
     \epsilon^{\text{u}} = -\ln\left\{Q\left[\rho_{\text{out},\pm}^{\text{u}}(\eta),\rho_{\text{out},\pm}^{\text{u}}(0)\right]\right\}.
\end{equation}

The trace in the $Q$ function then simplifies to:
\begin{widetext}
\begin{equation}
\mathrm{Tr}\left[\rho_{\text{out},\pm}^{\text{u}}(\eta)^s \rho_{\text{out},\pm}^{\text{u}}(0)^{1-s}\right] 
=\frac{1}{2} \mathrm{Tr}\left[ (\rho_{\text{out}}^{\text{QI}}(\eta) + \sigma^{\text{u}}(\eta))^s (\rho_{\text{out}}^{\text{QI}}(0) + \sigma^{\text{u}}(0))^{1-s} + (\rho_{\text{out}}^{\text{QI}}(\eta) - \sigma^{\text{u}}(\eta))^s (\rho_{\text{out}}^{\text{QI}}(0) - \sigma^{\text{u}}(0))^{1-s} \right].
\end{equation}
\end{widetext}

Note that $\rho_{\text{out}}^{\text{QI}}(\eta)-\sigma^{\text{u}}(\eta)$ for any $0\leq \eta\leq 1$ must be positive semi-definite, otherwise $\rho^{\text{u}}_{\text{out},\pm}$ will not be positive semi-definite and therefore will not be a physical density matrix.  By utilizing Lieb's Concavity Theorem~\cite{Lieb1973267,FAWZI2017240,Carlen2009TRACEIA} and Jensen's inequality, we can prove the following inequality
\begin{equation}   \label{eq:ineq}
    \epsilon^{\text{u}}\ge \epsilon^{\text{QI}}.
\end{equation}
The detailed proof of the above inequality can be found in Appendix \ref{app:a}.
We have demonstrated that
$\epsilon^{\text{u}}$ is always greater than or equal to $\epsilon^{\text{QI}}$. Therefore, we have \begin{equation}\label{eq:generic_advantage}
    e^{-\frac{1}{2}M\epsilon^{\text{u}}} \leq e^{-\frac{1}{2}M\epsilon^{\text{QI}}}.
\end{equation}
This proves that both ICO and PS-DE protocols can offer a rigorous advantage over standard QI, typically yielding a tighter upper bound on the error probability.

While Eq.~\eqref{eq:generic_advantage} establishes a general advantage, the magnitude of this advantage is bounded. To quantify this, we analyze the deviation
\begin{equation}
\begin{aligned}
g_s(\rho, \rho', \sigma, \sigma') & := \operatorname{Tr}(\rho^s \rho'^{1-s}) \\
& - \frac{1}{2} \operatorname{Tr}\left((\rho + \sigma)^s (\rho' + \sigma')^{1-s}\right) \\
& - \frac{1}{2} \operatorname{Tr}\left((\rho - \sigma)^s (\rho' - \sigma')^{1-s}\right),
\end{aligned}
\label{eq:g_s_def}
\end{equation}
which measures the reduction in the $Q_s$ function due to the interference terms. The following theorem provides an upper bound for this quantity (proven in Appendix~\ref{app:gpm}):
\begin{theorem}\label{th:gpm}
For density matrices \(\rho\), \(\rho'\) and Hermitian matrices \(\sigma\), \(\sigma'\) such that \(\rho \pm \sigma \geq 0\) and \(\rho' \pm \sigma' \geq 0\), there exist non-negative constants \(L_1(s), L_2(s)\) such that
\begin{equation}
\begin{aligned}
|g_s(\rho, \rho', \sigma, \sigma')| \leq& \mathcal{D} \left[ L_1(s) \lVert \sigma \rVert_{\text{spec}} (1 + \lVert \sigma' \rVert_{\text{spec}})^{1-s}\right.\\
+&\left. L_2(s) \lVert \sigma' \rVert_{\text{spec}} \right],
\end{aligned}
\end{equation}
where \(\mathcal{D}\) is the Hilbert space dimension and \(\lVert A \rVert_{\text{spec}}\) denotes the spectral norm of $A$, i.e., its the largest singular value. 
\end{theorem}
The proof of Theorem~\ref{th:gpm} is provided in Appendix~\ref{app:gpm}. We set \( Q_s(\rho, \rho') := \operatorname{Tr}(\rho^s \rho'^{1-s}) \) for \( 0 \leq s \leq 1 \), the $Q$ function is obtained by minimizing this function over \( s \). Let us now consider the function \( g_s \) defined in Eq.~\eqref{eq:g_s_def} with the identifications
\[
\rho = \rho_{\text{out}}^{\text{QI}}(\eta), \quad \rho' = \rho_{\text{out}}^{\text{QI}}(0), \quad \sigma = \sigma^{\text{u}}(\eta), \quad \sigma' = \sigma^{\text{u}}(0).
\]
For these choices, and noting the structure of the block-diagonalized state in Eq.~\eqref{eq:block_diagonal_form}, a direct calculation shows that \( g_s \) indeed measures the difference of the \( Q_s \) functions for the standard and superposed protocols before the minimization over \( s \):
\begin{equation}
g_s = Q_s\left[\rho_{\text{out}}^{\text{QI}}(\eta), \rho_{\text{out}}^{\text{QI}}(0)\right] - Q_s\left[\rho_{\text{out},\pm}^{\text{u}}(\eta), \rho_{\text{out},\pm}^{\text{u}}(0)\right].
\label{eq:g_s_identity}
\end{equation}

However, the optimal value \( s_{\text{u}}^* \) that minimizes \( Q_s[\rho_{\text{out},\pm}^{\text{u}}(\eta), \rho_{\text{out},\pm}^{\text{u}}(0)] \) is generally not the same as the value \( s_{\text{QI}}^* \) that minimizes \( Q_s[\rho_{\text{out}}^{\text{QI}}(\eta), \rho_{\text{out}}^{\text{QI}}(0)] \). Let us denote the minimal values as
\begin{equation}
\begin{aligned}
Q^{\text{QI}} &= \min_s Q_s[\rho_{\text{out}}^{\text{QI}}(\eta), \rho_{\text{out}}^{\text{QI}}(0)] = Q_{s_{\text{QI}}^*}^{\text{QI}}, \\
Q^{\text{u}} &= \min_s Q_s[\rho_{\text{out},\pm}^{\text{u}}(\eta), \rho_{\text{out},\pm}^{\text{u}}(0)] = Q_{s_{\text{u}}^*}^{\text{u}}.
\end{aligned}
\end{equation}
The difference in these minimal values is denoted by \( \Delta Q = Q^{\text{QI}} - Q^{\text{u}} \). Crucially, consider the value of \( g_s \) at \( s = s_{\text{u}}^* \), the minimizer for the superposition protocol:
\begin{equation}
g_{s_{\text{u}}^*} = Q_{s_{\text{u}}^*}^{\text{QI}} - Q_{s_{\text{u}}^*}^{\text{u}} = Q_{s_{\text{u}}^*}^{\text{QI}} - Q^{\text{u}}.
\end{equation}
Since \(Q^{\text{QI}} \leq Q_{s_{\text{u}}^*}^{\text{QI}} \) by definition of the minimum, it follows immediately that
\begin{equation}
\Delta Q = Q^{\text{QI}} - Q^{\text{u}} \leq Q_{s_{\text{u}}^*}^{\text{QI}} - Q^{\text{u}} = g_{s_{\text{u}}^*}.
\end{equation}
Applying Theorem~\ref{th:gpm} at the specific point \( s = s_{\text{u}}^* \) provides an upper bound for this term:
\begin{equation}
\Delta Q \leq g_{s_{\text{u}}^*} \leq \delta(s_{\text{u}}^*),
\end{equation}
where \( \delta(s) := \mathcal{D}( L_1(s) \lVert \sigma^{\text{u}}(\eta) \rVert_{\text{spec}} (1 + \lVert \sigma^{\text{u}}(0) \rVert_{\text{spec}})^{1-s} + L_2(s) \lVert \sigma^{\text{u}}(0) \rVert_{\text{spec}}) \).

This establishes a direct and rigorous upper bound on the potential advantage \( \Delta Q \) in terms of the bound for \( g_s \). Since the function \( \delta(s) \) is monotonic in the spectral norms \( \lVert \sigma^{\text{u}}(\eta) \rVert_{\text{spec}} \) and \( \lVert \sigma^{\text{u}}(0) \rVert_{\text{spec}} \), these norms indeed serve as key indicators of the protocol's potential performance gain.

This bound on the difference of the $Q_s$ functions, $\Delta Q \leq \delta(s_{\text{u}}^*)$, can be directly translated into a bound on the difference of the Chernoff exponents. Recall that the Chernoff exponents are defined by the minimization of the negative logarithm of $Q_s$:
\begin{equation}
\epsilon^{\text{QI}} = -\ln Q^{\text{QI}}, \quad \epsilon^{\text{u}} = -\ln Q^{\text{u}}.
\end{equation}
The difference in exponents is $\Delta \epsilon = \epsilon^{\text{u}} - \epsilon^{\text{QI}} = -\ln(Q^{\text{u}}) + \ln(Q^{\text{QI}}) = -\ln\left( Q^{\text{u}}/Q^{\text{QI}} \right) = -\ln\left(1 - \Delta Q/Q^{\text{QI}} \right)$.
Since the logarithm is a monotonically increasing function and $Q_{\text{min}}^{\text{u}} > 0$, the bound $\Delta Q \leq \delta(s_{\text{u}}^*)$ implies that the potential advantage in the Chernoff exponent is constrained by
\begin{equation}
\Delta \epsilon \leq -\ln\left(1 - \frac{\delta(s_{\text{u}}^*)}{Q^{\text{QI}}} \right).
\end{equation}
This establishes that a larger possible value of the upper bound $\delta$ permits a larger potential advantage $\Delta \epsilon$. 

\begin{figure}[t]
    \centering
    \includegraphics[width=\linewidth]{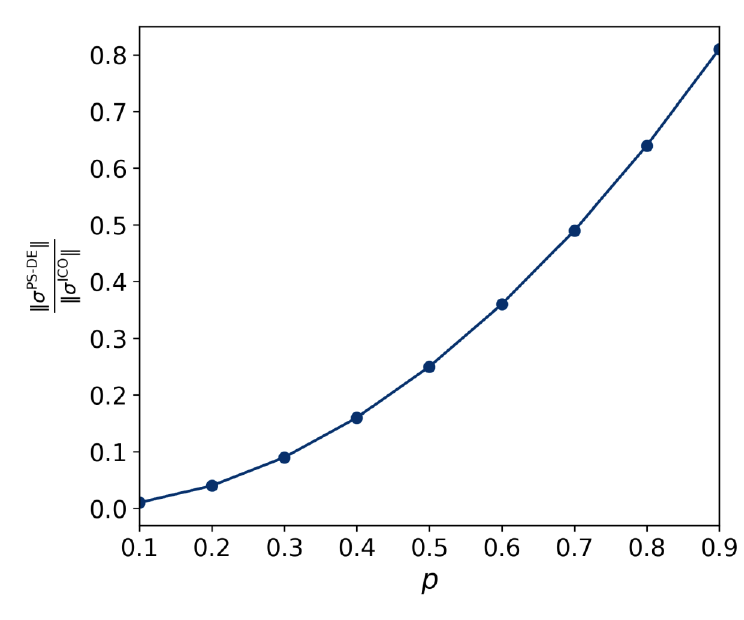}
    \caption{\justifying{\textbf{ICO Protocol Generates Stronger Quantum Interference Than PS-DE Protocol.} The ratio $||\sigma^{\text{PS-DE}}||{\text{spec}} / ||\sigma^{\text{ICO}}||{\text{spec}}$ is plotted against the survival probability $p$. The ratio increases from 0 to approximately 0.8, with its slope growing with $p$. The curves for different target reflectivities ($\eta = 0.01, 0.05, 0.1$) are nearly indistinguishable, indicating the ratio is robust against changes in $\eta$. These results demonstrate that the ICO protocol generates a consistently stronger quantum interference effect than the PS-DE protocol across a wide range of parameters.}}
    \label{fig:spectral_norm_ratio}
\end{figure}

Since $\delta(s)$ is monotonic in the spectral norms $\lVert \sigma^{\text{u}}(\eta) \rVert_{\text{spec}}$ and $\lVert \sigma^{\text{u}}(0) \rVert_{\text{spec}}$, this confirms the intuitive expectation that the performance advantage is fundamentally linked to the strength of the quantum interference effect.

The analytical results presented in this section, the guaranteed advantage Eq.~\eqref{eq:generic_advantage} and the bounded magnitude of this advantage, form the core theoretical contribution of this work. While these results confirm that any non-zero interference is beneficial, they do not by themselves predict a performance ranking between ICO and PS-DE. However, they provide a crucial theoretical lens for analysis: the established upper bound on the advantage is monotonic in the spectral norms $\lVert \sigma^{\text{u}}(\eta) \rVert_{\text{spec}}$ and $\lVert \sigma^{\text{u}}(0) \rVert_{\text{spec}}$.

This leads to a dual-layered implication. First, the spectral norm serves as an indicator of the \textit{potential} advantage, not its guarantee, as it governs an upper bound. Second, an analysis of the operator structure suggests that the ICO interference term $\sigma^{\text{ICO}}$ has a larger \textit{potential upper bound} for this spectral norm compared to $\sigma^{\text{PS-DE}}$, a consequence of its more complex summation over Kraus operators (Eq.~(\ref{eq:sigma_ICO})). This constitutes a theoretical hint that ICO might possess a greater capacity for error mitigation.

To move from this potential to a definitive performance comparison, and to explore the detailed dependence on system parameters such as the survival probability $p$ and the target reflectivity $\eta$, we must turn to detailed numerical simulations, which we undertake in Subsection~\ref{subsec:advantage}.

Having established the theoretical advantages of the channel superposition protocols, a critical question for their practical implementation is robustness against decoherence. The quantum advantage provided by these protocols relies on the coherence of the control qubit. Therefore, we now investigate the resilience of the ICO and PS-DE schemes to noise processes that directly affect the control qubit, i.e., phase and bit flips. This analysis will establish the resource-theoretic value of coherence within our framework and quantify the tolerance of the proposed protocols to experimental imperfections.

\section{Robustness of ICO and PS-DE Quantum Illumination under Control Qubit noise}\label{sec:resource}

In this section, we analyze the robustness of the ICO and PS-DE protocols to noise on the control qubit. We consider two fundamental types of noise: phase flip and bit flip. Our goal is to understand how these decoherence processes degrade the quantum interference effect and establish the resource-theoretic properties of the coherence in these specific channel superposition protocols.

Following the potential experimental setup described in Appendix~\ref{app:experimental_setup}, the quantum state of the system \textit{after} the probe modes have interacted with the target and noise channels but \textit{before} the final recombination at the receiver's controlled path swap operation, is given by:
\begin{equation}\label{eq:varrho}
\varrho_{\text{out}}^{\text{u}} = \frac{1}{2}
\begin{bmatrix}
\rho_{\text{out}, A_0B} ^{\text{QI}}\otimes |0\rangle_{A_1}\langle 0| & \sigma_{A_0A_1B}^{\text{u}} \\
\sigma_{A_1A_0B}^{\text{u}} & \rho_{\text{out},A_1B}^{\text{QI}} \otimes |0\rangle_{A_0}\langle 0|
\end{bmatrix}_c.
\end{equation}
Here, $\rho_{\text{out}, A_0B}^{\text{QI}}$ and $\rho_{\text{out},A_1B}^{\text{QI}}$ are the standard QI output states had the probe been sent only through path $A_0$ or $A_1$, respectively. The terms $\sigma_{A_0A_1B}^{\text{u}}$ encapsulate the quantum interference between these two possibilities.

We adopt the following nomenclature to clarify the experimental and mathematical description:
\begin{itemize}
    \item Physical Paths: The two distinct spatiotemporal trajectories of the probe are denoted by modes $A_0$ and $A_1$. The state $\varrho_{\text{out}}^{\text{u}}$ in Eq.~\eqref{eq:varrho} is expressed in this basis before the final recombination.
    \item Output Modes: The final measurement is performed on the interferometer's output ports. We denote the primary output mode (which contains the interference signal) as $A$, and the complementary, often unused, output mode as $\bar{A}$.
\end{itemize}
A detailed description of this setup and the corresponding unitary is provided in Appendix~\ref{app:experimental_setup}.

The unitary operation of the controlled path swap  $\mathcal{U}_\text{swap}^\dagger$ at the receiver performs the transformation $(A_0, A_1) \rightarrow (A, \bar{A})$, where $\mathcal{U}_{\text{swap}}$ is defined in Appendix ~\ref{app:experimental_setup}.  The final output state $\rho_{\text{out}}^{\text{u}}$, on which the measurement is performed, is obtained by applying $\mathcal{U}_\text{swap}^\dagger$ and subsequently tracing out the unused mode $\bar{A}$:
\begin{equation}
\rho_{\text{out}}^{\text{u}} = \Tr_{\bar{A}}\left( \mathcal{U}_{\text{swap}}^\dagger \, \varrho_{\text{out}}^{\text{u}} \, \mathcal{U}_{\text{swap}} \right) = \frac{1}{2}
\begin{bmatrix}
\rho_{\text{out}}^{\text{QI}} & \sigma^{\text{u}} \\
\sigma^{\text{u}} & \rho_{\text{out}}^{\text{QI}}
\end{bmatrix}_c.
\label{eq:final_output_state}
\end{equation}
This transformation maps the interference between paths $A_0$ and $A_1$ into an interference term $\sigma^{\text{u}}$ localized in the output mode $A$. We now analyze how noise on the control qubit affects this final state.

\subsection{Effects of Bit-Flip Noise}

The bit-flip channel on the control qubit is defined as $\mathcal{E}_c^{\text{BF}}(\rho_c) := \beta\rho_c + (1-\beta)X\rho_c X$, where $\beta \in [0,1]$ is the probability of no flip, and $X$ is the Pauli-$X$ operator. This noise decoheres the control qubit in the computational basis $\{ \ket{0}_c, \ket{1}_c \}$.

When bit-flip noise occurs \textit{before} the final recombination at the receiver, it acts on the state $\varrho_{\text{out}}^{\text{u}}$ defined in the physical path basis $(A_0, A_1)$ (in Eq.~\eqref{eq:varrho}). The resulting state is:
\begin{equation}
\mathcal{E}_c^{\text{BF}}(\varrho_{\text{out}}^{\text{u}}) = \beta \varrho_{\text{out}}^{\text{u}} + (1-\beta)\varrho_{\text{err}},
\end{equation}
where the error state $\varrho_{\text{err}}$ corresponds to the control qubit being flipped:
\begin{equation}
\varrho_{\text{err}} = \frac{1}{2}
\begin{bmatrix}
\rho_{\text{out},A_1B}^{\text{QI}} \otimes |0\rangle_{A_0}\langle 0| & \sigma_{A_1A_0B}^{\text{u}} \\
\sigma_{A_0A_1B}^{\text{u}\dagger} & \rho_{\text{out},A_0B}^{\text{QI}} \otimes |0\rangle_{A_1}\langle 0|
\end{bmatrix}_c.
\end{equation}
The subsequent recombination via the controlled-MZI and tracing out of mode $\bar{A}$ transforms these states into the final output basis:
\begin{align}
\rho_{\text{out}}^{\text{u}}(\beta) &= \beta \, \rho_{\text{out}}^{\text{u}} + (1-\beta) \, \rho_{\text{err}}, \quad \text{where} \\
\rho_{\text{err}} &= \Tr_{\bar{A}}\left( \mathcal{U}_{\text{swap}} \, \varrho_{\text{err}} \, \mathcal{U}_{\text{swap}}^\dagger \right).
\end{align}
The specific form of $\rho_{\text{err}}$ depends on the interferometer's operation. In the standard configuration where only output mode $A$ is measured, a flipped control qubit leads to the probe photon being routed to the wrong output port. This results in an effective loss channel on the measured mode $A$:
\begin{equation}
\rho_{\text{err}} = \left( \frac{1}{2} \begin{bmatrix} \rho_B & \sigma_B \\ \sigma_B & \rho_B \end{bmatrix}_c \right) \otimes |0\rangle_A\langle 0|,
\end{equation}
where $\rho_B = \operatorname{Tr}_{A}(\rho_{\text{out}}^{\text{QI}})$ and $\sigma_B = \operatorname{Tr}_{A}(\sigma^{\text{u}})$.

This bit-flip noise manifests as an effective loss channel on mode $A$: $\mathcal{E}(\rho_A) = \beta \rho_A + (1-\beta)|0\rangle_A\langle 0|$, thereby degrading performance. To mitigate this, we propose a robust measurement scheme: jointly measuring both output modes $A$ and $\bar{A}$ (see Fig.~\ref{fig:robustness}). This strategy preserves the information from both interferometer ports. The final measurement is then performed on the combined state $\mathcal{U}_{\text{swap}}^\dagger \, \varrho_{\text{out}}^{\text{u}} \, \mathcal{U}_{\text{swap}}$ in the $(A, \bar{A})$ basis, rather than on $\rho_{\text{out}}^{\text{u}} = \Tr_{\bar{A}}(\cdots)$. This ensures that the quantum interference information, present in both $\varrho_{\text{out}}^{\text{u}}$ and $\varrho_{\text{err}}$, is fully retained in the measurement statistics, thereby recovering the advantage over standard QI even under bit-flip noise.

\subsection{Effects of Phase-Flip Noise}

The phase-flip channel on the control qubit is defined as $\mathcal{E}_c^{\text{PF}}(\rho_c) := \beta \rho_c + (1-\beta) Z \rho_c Z$, where $\beta \in [0,1]$ is the probability of no phase flip, and $Z$ is the Pauli-$Z$ operator. This noise decoheres the control qubit in the $\{ \ket{+}_c, \ket{-}_c \}$ basis.

We model phase-flip noise as acting \textit{after} the final recombination and measurement basis transformation. This is a realistic model for noise that degrades the phase coherence of the control qubit at the stage of final measurement or due to imperfections in the interferometer itself. Under this noise model, the final output state $\rho_{\text{out}}^{\text{u}}$ (in Eq.~\eqref{eq:final_output_state}) becomes:
\begin{equation}
\rho_{\text{out}}^{\text{u}}(\beta) = \mathcal{E}_c^{\text{PF}}(\rho_{\text{out}}^{\text{u}}) = \frac{1}{2}\left[
\begin{matrix}
\rho_{\text{out}}^{\text{QI}} & (2\beta-1)\sigma^{\text{u}} \\
(2\beta-1)\sigma^{\text{u}} & \rho_{\text{out}}^{\text{QI}}
\end{matrix}
\right]_c.
\label{eq:output_pf}
\end{equation}
To simplify the analysis, we define a \textit{decoherence coefficient} $\gamma = 2\beta - 1$, which quantifies the residual coherence, with $\gamma = 1$ ($\beta=1$) indicating no noise and $\gamma = 0$ ($\beta=1/2$) indicating complete dephasing. The parameter $\gamma$ scales the off-diagonal interference terms, directly reflecting the strength of the remaining quantum interference effect.

Transforming this state into the eigenbasis $\{\ket{+}_c, \ket{-}_c\}$ of the control qubit yields a block-diagonal form:
\begin{equation}
\rho_{\text{out},\pm}^{\text{u}}(\gamma) = \frac{1}{2}\left[
\begin{matrix}
\rho_{\text{out}}^{\text{QI}} + \gamma\sigma^{\text{u}} & 0 \\
0 & \rho_{\text{out}}^{\text{QI}} - \gamma\sigma^{\text{u}}
\end{matrix}
\right]_c.
\label{eq:diag_phaseflip}
\end{equation}
This decomposition clearly shows that the phase flip noise attenuates the quantum interference by the factor $\gamma$.

The impact of this decoherence on the protocol's performance is quantified by the $\gamma$-dependent Chernoff exponent $\epsilon^{\text{u}}(\gamma) = -\ln\left( \min_{s} Q_s[\rho_{\text{out},\pm}^{\text{u}}(\gamma), \rho_{\text{out},\pm}^{\text{u}}(0)] \right)$. To analyze this, we define the function $f(\rho, \rho') = \Tr[\rho^s \rho'^{1-s}]$ as in Appendix~\ref{app:a}, which is jointly concave for $s \in [0,1]$. The relevant quantity for the QCB is:
\begin{equation}
\begin{aligned}
F_s(\gamma) &:= \frac{1}{2} \left[ f(\rho_{\text{out}}^{\text{QI}} + \gamma\sigma^{\text{u}}, \rho_{\text{out}}^{\text{QI}} + \gamma\sigma^{\text{u}}) \right.\\
&+ \left. f(\rho_{\text{out}}^{\text{QI}} - \gamma\sigma^{\text{u}}, \rho_{\text{out}}^{\text{QI}} - \gamma\sigma^{\text{u}}) \right],
\end{aligned}
\end{equation}
where we have suppressed the $\eta$ and $0$ arguments for the states for clarity, and note that $\sigma^{\text{u}}$ is evaluated at the same $\eta$ as its accompanying $\rho_{\text{out}}^{\text{QI}}$.

The monotonicity of the performance with respect to the resource $\gamma$ is established by the following theorem (proven in Appendix~\ref{app:b}).
\begin{theorem}\label{th:1}
Let $\rho$, $\rho'$ be density matrices and $\sigma$, $\sigma'$ be Hermitian matrices such that $\rho \pm \sigma \geq 0$ and $\rho' \pm \sigma' \geq 0$. For $s \in [0,1]$, the function
\begin{equation}\label{eq:fgamma}
F_s(\gamma) = \frac{1}{2} \left[ f(\rho + \gamma\sigma, \rho' + \gamma\sigma') + f(\rho - \gamma\sigma, \rho' - \gamma\sigma') \right]
\end{equation}
is monotonically decreasing in $\gamma$ for $\gamma \in [0,1]$.
\end{theorem}

Since $F_s(\gamma)$ is an even function, we restrict our analysis to $\gamma \in [0,1]$. The monotonicity of $F_s(\gamma)$ implies that the minimized value $F_{\text{min}}(\gamma) = \min_{s \in [0,1]} F_s(\gamma)$ is also monotonically decreasing in $\gamma$. Consequently, the Chernoff exponent $\epsilon^{\text{u}}(\gamma) = -\ln F_{\text{min}}(\gamma)$ is monotonically increasing in $\gamma$:
\begin{equation}
\epsilon^{\text{u}}(\gamma_1) \leq \epsilon^{\text{u}}(\gamma_2) \quad \text{for} \quad 0 \leq \gamma_1 < \gamma_2 \leq 1.
\end{equation}
This establishes the decoherence coefficient $\gamma$ as a quantifier of the useful resource: a larger $|\gamma|$ signifies a greater amount of preserved coherence between the two channels, which in turn enables a lower error probability bound.

The parameter $\gamma$ is directly related to the fidelity between the noisy control qubit state and the ideal pure state. For the phase-flip channel, the fidelity $\mathcal{F}$ between $\mathcal{E}_c^{\text{PF}}(\ket{+}\bra{+})$ and $\ket{+}\bra{+}$ is $\mathcal{F} = (1 + \gamma)/2$.

This framework naturally admits a resource-theoretic interpretation:
\begin{itemize}
    \item[•] \textbf{Monotonicity:} The performance metric $\epsilon^{\text{u}}(\gamma)$ is a monotonic function of the resource $|\gamma|$.
    \item[•] \textbf{Free State:} The state with $\gamma = 0$ (complete dephasing) is the free state, corresponding to standard quantum illumination without any superposition advantage.
    \item[•] \textbf{Free Operations:} Operations that do not increase $|\gamma|$ are considered free operations within this specific noise model.
\end{itemize}

The combined analysis of bit-flip and phase-flip noise provides a comprehensive picture of the robustness of the ICO and PS-DE protocols. General forms of control-qubit noise can be approximated as sequences or convex combinations of these two fundamental channels. Our analysis reveals distinct impacts and mitigation strategies for each type.

Bit-flip noise primarily degrades performance by causing information loss through the discarding of the $\bar{A}$ mode in the standard measurement setup. As proposed, this can be actively mitigated by a robust measurement strategy that jointly measures both output modes $A$ and $\bar{A}$, thereby recovering the information that would otherwise be lost. In contrast, phase-flip noise reduces the performance of these protocols passively and continuously by attenuating the strength of the quantum interference effect, as quantified by the decoherence coefficient $\gamma$. Crucially, the theory based on Lieb's concavity theorem (Appendix~\ref{app:a}) guarantees that any non-zero residual coherence ($\gamma > 0$) still provides an advantage over the standard quantum illumination protocol, i.e., $\epsilon^{\text{u}}(\gamma) \geq \epsilon^{\text{QI}}$ for $\gamma > 0$. The performance thus gracefully decreases as $\gamma$ approaches zero.

In summary, the ICO and PS-DE protocols exhibit inherent robustness to realistic noise on the control qubit. Bit-flip errors can be circumvented through an adapted measurement strategy, while phase-flip errors lead to a graceful degradation of performance, maintaining a quantum advantage for any non-vanishing level of coherence.

\begin{figure*}[t]
    \centering
    \includegraphics[width=\linewidth]{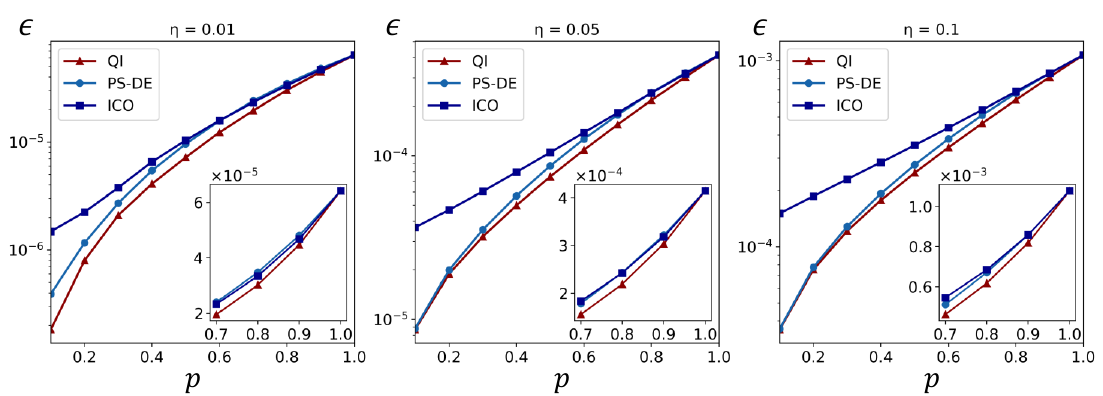}
    \caption{\textbf{Numerical simulation results showing the Chernoff exponent $\epsilon$ for QI, PS-DE, and ICO protocols as a function of survival probability $p$.} 
    \justifying{Results are obtained at fixed truncation dimension $\mathcal{D}=10$ and for reflectivities $\eta=0.01, \eta=0.05$, and $\eta=0.1$. The ICO protocol exhibits superior performance for $p \lesssim 0.6$--$0.7$, beyond which the PS-DE protocol performs comparably.}}
    \label{fig:simulation_p}
\end{figure*}

\begin{figure*}[t]
    \centering
    \includegraphics[width=\linewidth]{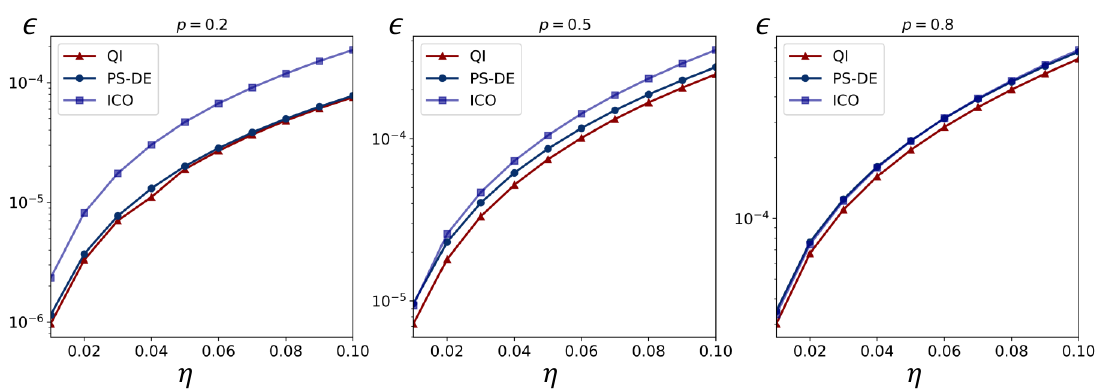}    \captionsetup{justification=justified, singlelinecheck=false}
    \caption{\textbf{Dependence of $\epsilon$ on target reflectivity $\eta$ for QI, ICO, and PS-DE protocols.}
    \justifying{This figure illustrates how the Chernoff exponent $\epsilon$ varies with the target reflectivity $\eta$ across different loss channel probabilities $p=0.2, p=0.5$, and $p=0.8$, with truncation dimension fixed at $\mathcal{D}=10$. The results demonstrate that $\epsilon$ increases monotonically with $\eta$ for all three protocols (QI, ICO, and PS-DE).}}
    \label{fig:simulation_eta}
\end{figure*}

\section{Numerical simulation}
\label{sec:simulation}

The preceding theoretical analysis established the foundation and predicted a quantum advantage for the ICO and PS-DE protocols utilizing channel superposition.
This section presents a numerical analysis to validate these theoretical findings and to quantitatively compare the performance of the ICO and PS-DE protocols under various conditions of photon loss probability \( p \) and target reflectivity \( \eta \). Our simulations have three primary objectives: to verify the convergence of our model with respect to the Hilbert space truncation dimension \(\mathcal{D}\), to compare the innate interference strength generated by the ICO and PS-DE protocols, and to numerically compute the quantum Chernoff bound to assess the resulting advantage in target detection performance.

\subsection{Hilbert Space Truncation and Convergence}\label{sec:setup}

The analytical expressions for the output states and the QCB involve operators acting on infinite-dimensional Hilbert spaces. For numerical simulation, a finite-dimensional truncation is necessary. This subsection details our initial state preparation and justifies the choice of truncation dimension $\mathcal{D}$ by demonstrating the convergence of the key quantity $\epsilon^{\text{u}}$.

The initial bipartite optical state is prepared as a photon-subtracted two-mode squeezed state, a non-Gaussian state known to enhance QI performance in noisy environments. Its wavefunction is given by:
\begin{equation}
    \ket{\psi}_{AB}:= \sum_n c_n \ket{n}_A\otimes \ket{n}_B,
\end{equation}
where the coefficients $c_n$ are defined as $c_n = (1-\lambda^2)^{3/2}(n+1)\lambda^n/\sqrt{1+\lambda^2}$, $0 \leq \lambda \leq 1$ is the squeezing parameter, and the factor $(n+1)$ is the signature of photon subtraction. This state represents an entangled state that can be used for quantum illumination purposes. The mean photon number in the transmitted mode $A$ is $N_t = \langle a_A^\dagger a_A \rangle$.

The target and loss channels are implemented using the Kraus operator formalism detailed in Sec.~\ref{sec:quantum_illumination} and Sec.~\ref{sec:ICO}. The decisive parameter for numerical feasibility is the truncation dimension $\mathcal{D}$, which defines the maximum photon number $n_{\text{max}} = \mathcal{D} - 1$ considered in the simulation. The choice of $\mathcal{D}$ must be large enough to ensure that the calculated Chernoff exponent $\epsilon^{\text{u}}$ converges to its true value within the relevant parameter regime.

To determine a sufficient value for $\mathcal{D}$, we performed convergence tests. Fig.~\ref{fig:dimension} illustrates the dependence of $\epsilon^{\text{QI}}$, $\epsilon^{\text{PS-DE}}$, and $\epsilon^{\text{ICO}}$ on the truncated dimension $\mathcal{D}$ for a representative set of parameters: target reflectivity $\eta = 0.01$, environmental thermal noise mean photon number $N=0.5$, loss channel survival probability $p = 0.9$, and a low transmitted photon number $N_t \approx 0.01$. The results indicate that for $\mathcal{D} \geq 10$, all three exponents stabilize, confirming that the numerical results have converged. Based on this analysis, we set the truncation dimension to $\mathcal{D} = 10$ for all subsequent simulations. This choice ensures computational tractability while guaranteeing that our results are not artifacts of the finite-dimensional approximation.

\subsection{Comparative Analysis of Interference Strength}\label{subsec:comparison}

The superior robustness of the ICO protocol is fundamentally rooted in the greater strength of the quantum interference it generates, as quantified by the spectral norm of the off-diagonal term $\lVert \sigma^{\text{u}} \rVert_{\text{spec}}$ in the output state. As established in Section~\ref{sec:QCB}, this norm serves as a key indicator of the potential advantage in the Chernoff exponent.

Our numerical simulations confirm this theoretical insight. Fig.~\ref{fig:spectral_norm_ratio} directly compares the interference strength of the two protocols by plotting the ratio $\lVert \sigma^{\text{PS-DE}} \rVert_{\text{spec}} / \lVert \sigma^{\text{ICO}} \rVert_{\text{spec}}$ across the range of survival probability $p$ from $0.1$ to $0.9$. The ratio increases from zero but remains consistently below unity, reaching a maximum of only approximately 0.8 at $p=0.9$. This demonstrates that the ICO protocol maintains a stronger interference effect, with the relative advantage being most pronounced in the high-loss regime (low $p$). Furthermore, this ratio exhibits negligible dependence on the target reflectivity $\eta$, as evidenced by the nearly identical curves for $\eta = 0.01, 0.05,$ and $0.1$. This robustness to $\eta$ underscores the fundamental nature of the advantage, which stems from the protocol architecture itself rather than specific target parameters.

\subsection{Performance Advantage and Resource Resilience of the ICO Protocol}\label{subsec:advantage}

Having established that the ICO protocol generates a fundamentally stronger innate interference effect than the PS-DE protocol, we now present a comprehensive numerical analysis of its performance advantage over standard quantum illumination. Furthermore, we examine the resilience of this advantage to noise on the control qubit, directly connecting it to the resource-theoretic measure of coherence, $\gamma$, introduced in Section~\ref{sec:resource}.

Fig.~\ref{fig:simulation_p} reveals that the advantage of the ICO/PS-DE protocol, $\epsilon^{\text{u}} > \epsilon^{\text{QI}}$, is most pronounced in high-loss scenarios (low $p$). This is the regime where preserving the non-classical correlations between the probe and the ancilla is most challenging, and thus the error-mitigating effect of the ICO/PS-DE provides the greatest relative benefit. As the loss decreases (high $p$), the advantage diminishes, as both protocols eventually operate in a regime where photon loss is no longer the dominant source of error.

Fig.~\ref{fig:simulation_eta} demonstrates that the ICO protocol maintains its superior performance across different target reflectivities $\eta$. The absolute advantage is largest for very low-reflectivity targets, which is precisely the regime where quantum illumination is most valuable compared to classical alternatives.

\begin{figure}[t]
    \centering
    \includegraphics[width=\linewidth]{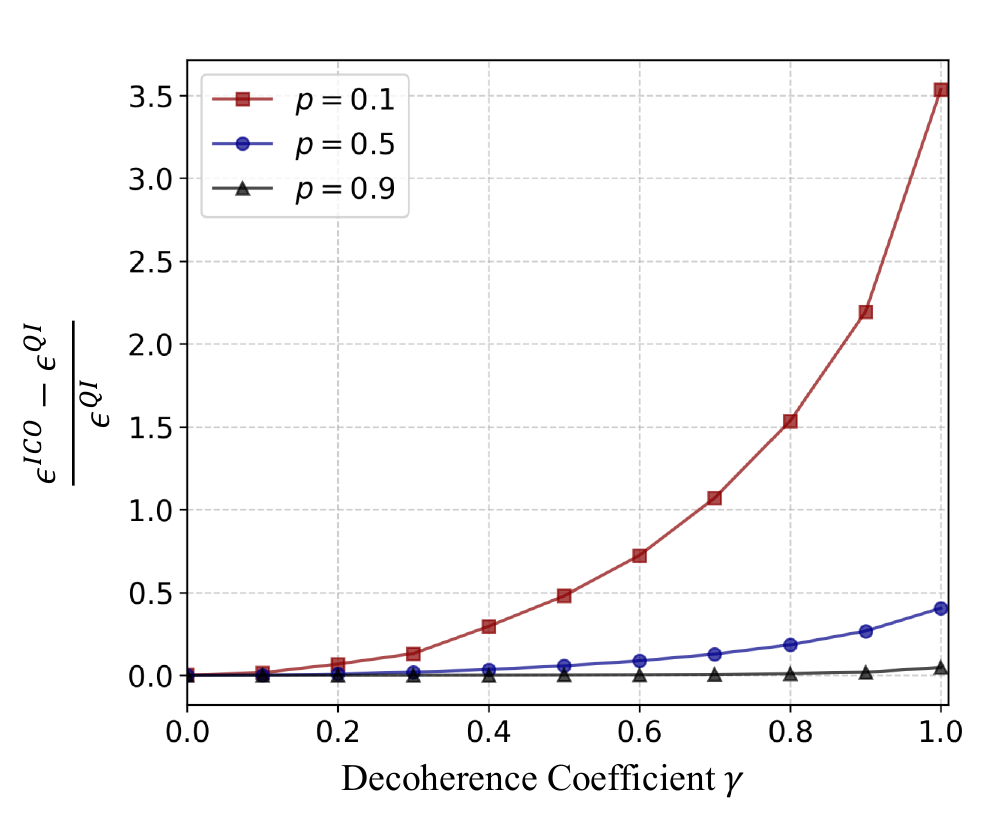}    \captionsetup{justification=justified, singlelinecheck=false}
    \caption{\textbf{Resource monotonicity in ICO.} 
    \justifying{Variation of $(\epsilon^{ICO}-\epsilon^{QI})/\epsilon^{QI}$ as a function of decoherence coefficient $\gamma$, a parameter quantifying the strength of ICO resources. The reflectivity is set as $\eta=0.05$.}}
    \label{fig:performance_comparison}
\end{figure}

The performance advantage is underpinned by the availability of the quantum coherence resource quantified by $|\gamma|$. Fig.~\ref{fig:performance_comparison} illustrates this relationship by plotting the relative improvement $(\epsilon^{\text{ICO}} - \epsilon^{\text{QI}})/\epsilon^{\text{QI}}$ as a function of the decoherence coefficient $|\gamma|$. This figure validates the theoretical monotonicity proven in Section~\ref{sec:resource}: the performance of the ICO protocol degrades gracefully as the coherence resource $|\gamma|$ is depleted by phase-flip noise. Crucially, any non-zero amount of residual coherence ($\gamma > 0$) translates into a non-zero performance advantage over the standard QI protocol. This robust, monotonic relationship firmly establishes $|\gamma|$ as a quantifiable resource for enhanced target detection in lossy and noisy environments.

In conclusion, the numerical simulations confirm that the ICO protocol delivers a significant and robust performance advantage. It outperforms both the standard protocol and the more mathematically general PS-DE approach by generating a stronger interference effect that is more resilient to photon loss. This advantage is maintained even under considerable decoherence of the control qubit, fading only when the quantum coherence resource is completely exhausted ($\gamma = 0$).

\section{ICO vs. coherent control of the environment}\label{sec:origin}

In previous sections, we have theoretically established the quantum advantages of two channel superposition schemes for quantum illumination, the shared-environment approach (ICO) and the environment-independent approach (PS-DE), in mitigating photon loss.
In this section, we investigate a third type of protocol. The protocol supposed coherent control of the environment systems. The protocol was introduced in discussions concerning whether ICO performance advantages are shared with other schemes of coherent superpositions of channels~\cite{Abbott2020communication,PhysRevA.107.062208,PhysRevA.99.062317}.
More specifically, in reference~\cite{Abbott2020communication}, the authors introduced a coherent control mechanism that operates not only on the system but also on the environment while maintaining definite causal order (DCO). Their work demonstrated that for communication tasks through zero-capacity depolarizing channels, such coherent quantum channel control can achieve performance advantages equivalent to those offered by ICO.

In the context of our ICO scheme, we recognize that similar questions may arise regarding the origin of its performance advantage, even though our primary objective remains the development of an enhanced quantum illumination protocol.  
Through detailed analysis, we demonstrate that when environmental manipulation is permitted, an equivalent advantage can be achieved through a DCO implementation. 

\begin{figure}
    \centering
    \includegraphics[width=\linewidth]{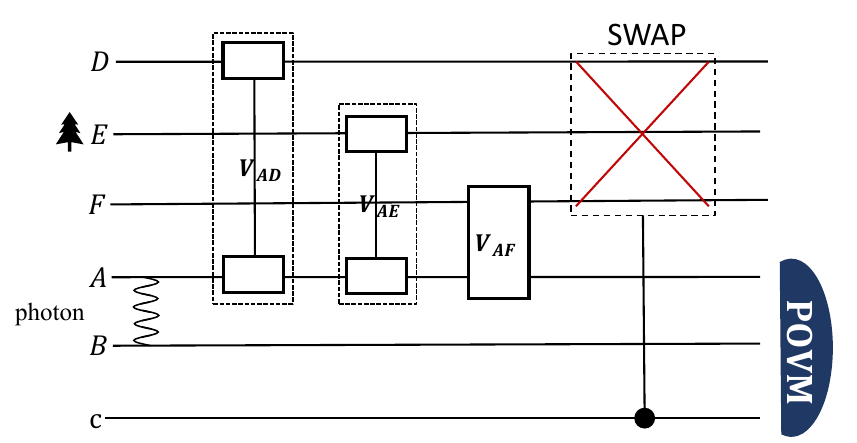}
    \caption{\textbf{Definite causal order implementation with environmental control.} \justifying{The setup consists of a control qubit (c), the probe photon interacting sequentially with three channel units $(V_{AD}, V_{AE}, V_{AF})$, and a SWAP gate acting on two corresponding auxiliary modes.}}
    \label{fig:DCO}
\end{figure}

The DCO protocol is illustrated in Fig.~\ref{fig:DCO}. The probe mode $A$ sequentially passes through the loss channel $\mathcal{E}^{(D)}$, the probe channel $\mathcal{E}^{(E)}$, and a second loss channel $\mathcal{E}^{(F)}$. 
These channels can be extended to unitary operations $V_{AD}$, $V_{AE}$, $V_{AF}$, supplemented by corresponding environment spaces $D$, $E$, and $F$. 

After applying these unitaries and the controlled-SWAP transformation, $\ket{0}_c\bra{0} U_{\text{swap}} + \ket{1}_c\bra{1} \otimes \mathbb{I}$, we trace out all environmental degrees of freedom. The resulting quantum state components, expressed in the environmental basis $\ket{i}_{D}\ket{jk}_E\ket{l}_F$, can be written as:
\begin{equation}
\begin{aligned}
&\bra{ijkl}_{DEF} \ket{\psi}_{\text{out}} \\
&= \frac{1}{\sqrt{2}} \left[ \ket{0}_c \otimes \bra{ijkl}_{DEF} U_{\text{swap}} V_{AF} V_{AE} V_{AD} \ket{\psi}_{AB} \otimes \ket{\psi}_{DEF} \right.\\
&\left.+ \ket{1}_c \otimes \bra{ijkl}_{DEF} V_{AF} V_{AE} V_{AD} \ket{\psi}_{AB} \otimes \ket{\psi}_{DEF} \right],
\end{aligned}
\end{equation}
where $U_{\text{swap}}$ is defined as:
\begin{equation}
U_{\text{swap}} = \sum_{ij} \ket{ij}_{DF} \bra{ji} + \ket{ji}_{DF} \bra{ij}.
\end{equation}
The inner product expression can therefore be reformulated as:
\begin{equation}
\begin{split}
&\bra{i}_D\bra{jk}_E\bra{l}_F \ket{\psi}_{\text{out}} \\
&= \frac{1}{\sqrt{2}} \left[K_i^{(D)} K_{jk}^{(E)} K_l^{(F)} \ket{\psi}_{AB} + K_l^{(D)} K_{jk}^{(E)} K_i^{(F)} \ket{\psi}_{AB}\right],
\end{split}
\end{equation}
Given that in our setup the Kraus operators satisfy $K_i^{(D)} = K_i^{(F)} = K_i$, this expression simplifies to:
\begin{equation}
\begin{split}
&\bra{ijkl}_{DEF} \ket{\psi}_{\text{out}} \\&= \frac{1}{\sqrt{2}}\left[ K_i K_{jk} K_l \ket{\psi}_{AB} + K_l K_{jk} K_i \ket{\psi}_{AB}\right],
\end{split}
\end{equation}
Consequently, we obtain the output state density matrix, expressed as a block matrix in the control qubit basis:
\begin{equation}\label{eq:DCO_output_state}
\rho_{\text{out}}^\text{DCO} =\frac{1}{2} \begin{bmatrix}
\rho^\text{QI} & \sigma^\text{DCO} \\
\sigma^\text{DCO} & \rho^\text{QI}
\end{bmatrix},
\end{equation}
where the interference term is given by:
\begin{equation}\label{eq:sigma_DCO}
\sigma^\text{DCO} = \frac{1}{2}\sum_{ijkl} K_i K_{jk} K_{l} \rho_{\text{in}} K_i^\dagger K_{jk}^\dagger K_l^\dagger.
\end{equation}
Comparing Eq.~\eqref{eq:DCO_output_state}
with Eq.~\eqref{eq:unified_output_structure}
and comparing Eq.~\eqref{eq:sigma_DCO} with Eq.~\eqref{eq:sigma_ICO}, we see that they are identical. This derivation thus establishes that the ICO quantum illumination scheme and a DCO scheme augmented by controlled environmental SWAP operations yield the same final state and thus the same performance. The ICO scheme does not require the SWAP gate and is in that sense less costly to implement.

\section{Conclusion}
\label{sec:conclusion}

In this work, we have investigated two specific channel superposition protocols---ICO and PS-DE---for quantum illumination, demonstrating their effectiveness in mitigating the detrimental effects of photon loss, a critical challenge in practical quantum sensing. 

Our analytical results form a solid theoretical foundation. We proved that any non-zero quantum interference, enabled by a coherent control qubit, guarantees a performance advantage over standard quantum illumination by yielding a tighter upper bound on the error probability. Furthermore, we derived that the magnitude of this potential advantage is fundamentally governed by the spectral norm of the interference term, $\lVert \sigma \rVert_{\text{spec}}$, in the protocol's output state.

A key insight from our work is the stark qualitative difference between ICO and PS-DE protocols, rooted in their use of environmental modes. By applying operations in a superposition of temporal orders to a shared environment, the ICO protocol generates a complex interference term $\sigma^{\text{ICO}}$ that is inherently more resilient to photon loss within the channels. In contrast, the PS-DE protocol interferes with two distinct environmental interactions, resulting in a simpler term $\sigma^{\text{PS-DE}} = p^2\mathcal{E}_\eta(\rho_{AB})$ that is more susceptible to decoherence.

Our numerical simulations validated these theoretical predictions. We first confirmed that the interference strength, as measured by the spectral norm, is consistently stronger for ICO than for PS-DE across all loss parameters (Fig.~\ref{fig:spectral_norm_ratio}). This directly translated into a superior performance advantage for ICO, evidenced by a higher quantum Chernoff exponent $\epsilon$,
particularly in the high-loss, low-reflectivity regime where quantum illumination is most valuable (Figs.~\ref{fig:simulation_p} and ~\ref{fig:simulation_eta}). Finally, we embedded this advantage within a resource-theoretic framework, showing that the ICO protocol's performance benefits degrade gracefully with control qubit noise and can be made robust against bit-flip errors through an adapted measurement strategy.

In summary, we have shown that indefinite causal order is not merely a conceptual curiosity but a definable, quantifiable resource that provides a tangible and robust advantage in the practical metrological task of quantum illumination. Looking ahead, this work opens several promising avenues, including the experimental implementation of an ICO-based quantum illumination protocol using integrated photonics, the extension to the superposition of more than two channels, and the application of these principles to other quantum sensing paradigms such as spectroscopy or communication through turbulent channels.

\section*{acknowledgments}
We gratefully acknowledge discussions with Fei Meng. We acknowledge support from the National Natural Science Foundation of China (Grants No. 12050410246, No.1200509, No.12050410245) and the City University of Hong Kong (Project No. 9610623).

%

\onecolumngrid
\newpage 
\begin{appendix}
\appendix 
\begin{figure}[H]
    \centering
    \includegraphics[width=\linewidth]{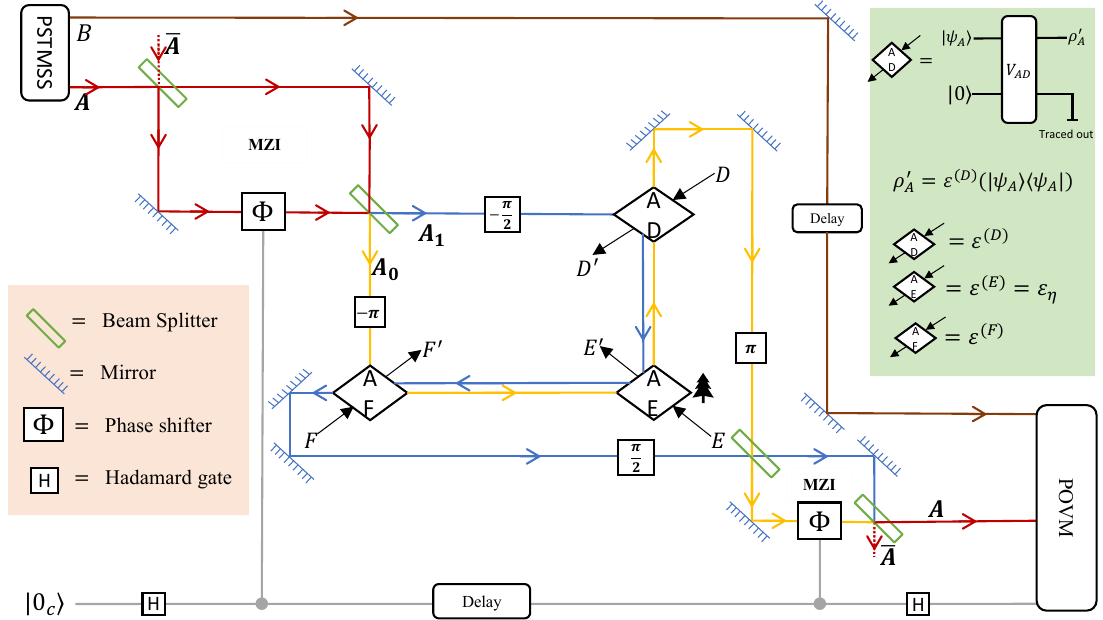}
    \caption{\textbf{Schematic of an MZI with Control System.} \justifying{This figure illustrates the setup of an MZI with a control system that modulates the phase shift based on the state of a control qubit. The control system adjusts the phase $\Phi$ in the interferometer path, enabling conditional phase shifts that are dependent on the control qubit's state. 
    The beam splitters and mirrors direct the light paths.
    The initial bipartite optical state is prepared in a photon-subtracted two-mode squeezed state (PSTMSS).
    The diagram also shows the interaction with auxiliary modes $D, E$, and $F$ through unitary operations $V_{AD}$, $V_{AE}$ and $V_{AF}$. 
    Light traverses various paths, undergoes coherent recombination at the end of the MZI, and is subsequently subjected to the overall measurement. The schematic includes components such as beam splitters, mirrors, phase shifters, and Hadamard gates, which facilitate the precise manipulation and measurement of quantum states.}}
    \label{fig:MZ-interferometer}
\end{figure}
\section{Potential Experimental Configuration}\label{app:experimental_setup}
This appendix outlines a potential experimental implementation of a quantum illumination protocol utilizing a control qubit to achieve channel superposition. The core idea is to use a Mach-Zehnder interferometer (MZI) whose phase shift is controlled by the state of a quantum bit to route a photonic probe through different spatiotemporal paths. 
To avoid redundancy and streamline the presentation, we focus on the ICO protocol as illustrated in Fig.~\ref{fig:MZ-interferometer}, since the PS-DE protocol follows an identical structural framework and thus requires no separate description.
As shown in Fig.~\ref{fig:MZ-interferometer}, our proposed experimental setup is based on a controlled Mach–Zehnder interferometer (MZI). To simulate the channels along the path, we introduce three auxiliary modes $D$, $E$, $F$ which interact unitarily with the transmitted modes $A_0$ or $A_1$ via the respective unitary operators $V_{AD}$, $V_{AE}$, $V_{AF}$.

\subsection{Setup and State Preparation}
The initial bipartite optical state is a photon-subtracted two-mode squeezed state (PSTMSS), shared between a probe mode $A$ and an ancilla mode $B$. A control qubit is prepared in $\ket{+}_c = (\ket{0}_c + \ket{1}_c)/\sqrt{2}$.
The controlled path-swap operator $\mathcal{U}_{\text{swap}}$ is implemented using a Mach-Zehnder interferometer (MZI) with a phase shifter controlled by the qubit state, followed by a fixed phase compensation network $\mathcal{P}$. 

The operational principle of the $\mathcal{U}_\text{swap}$ is as follows: as illustrated in Fig.~\ref{fig:MZ-interferometer}, it consists of two beam splitters, two mirrors, and a phase shifter.
In the Heisenberg picture, the action of $\mathcal{U}_\text{swap}$ is:
\begin{equation}\label{eq:uswap}
    \mathcal{U}^\dagger_\text{swap} \left( \mathbb{I}_c \otimes \begin{bmatrix} a_A \\ a_{\bar{A}} \end{bmatrix} \right) \mathcal{U}_\text{swap}
    = \ket{0}_c\bra{0} \otimes \begin{bmatrix}  a_{A_0} \\ a_{A_1} \end{bmatrix} 
    + \ket{1}_c\bra{1} \otimes \begin{bmatrix}  a_{A_1} \\  a_{A_0} \end{bmatrix},
\end{equation}
where \(\bar{A}\) denotes the vacuum auxiliary mode orthogonal to mode \( A \) (with no incident light). Following we provide a detailed description of the construction of the $\mathcal{U}_\text{swap}$ operator.

The beam splitter is characterized by the unitary operator
\begin{equation}
    S(\theta) = \exp\left[ i\theta \left( a_A^\dagger a_{\bar{A}} + a_A a_{\bar{A}}^\dagger \right) \right].
\end{equation}
The corresponding transform matrix under the Heisenberg transformation can be obtained by 
\begin{equation}
    S(\theta)^\dagger \begin{bmatrix}
        a_A\\a_{\bar{A}}
    \end{bmatrix}S(\theta)=\left[\begin{matrix}\cos(\theta) & i\sin(\theta)\\i\sin(\theta) &\cos(\theta)\end{matrix}\right]\begin{bmatrix}
        a_A\\a_{\bar{A}}
    \end{bmatrix}.
\end{equation}
The phase shifter is described by the operator
\begin{equation}
    P(\Phi) = \exp\left( i\Phi a^\dagger a \right),
\end{equation}
with \( a^\dagger \) being the creation operator for the mode, and the action of the phase shifter being $P(\Phi)^\dagger a P(\Phi)=e^{i\Phi}a$. The overall Heisenberg transformation of the MZI can be expressed through the transfer matrix:
\begin{equation}
    \begin{bmatrix} a_{A_0}(\Phi) \\ a_{A_1}(\Phi) \end{bmatrix}
    = U_\text{MZI}(\Phi)^\dagger \begin{bmatrix} a_A \\ a_{\bar{A}} \end{bmatrix} U_\text{MZI}(\Phi)=\frac{1}{\sqrt{2}}\left[\begin{matrix}1& i\\i & 1\end{matrix}\right]\left[\begin{matrix}1 & \\ &e^{i\Phi}\end{matrix}\right]\frac{1}{\sqrt{2}}\left[\begin{matrix}1& i\\i & 1\end{matrix}\right]\begin{bmatrix} a_A \\ a_{\bar{A}} \end{bmatrix}=ie^{i\Phi/2}\begin{bmatrix}-\sin(\Phi/2)a_A +\cos(\Phi/2)a_{\bar{A}}\\
    \cos(\Phi/2)a_A +\sin(\Phi/2)a_{\bar{A}}
    \end{bmatrix},
\end{equation}
where $ U_\text{MZI}(\Phi) = S(\pi/4) P_{\bar{A}}(\Phi) S(\pi/4)$ represents the composite unitary operation. Thus, we have $a_{A_0}(0) = ia_{\bar{A}}$, $a_{A_1}(0)=i a_{A}$, and $a_{A_0}(\pi)=-a_{A}$, $a_{A_1}(\pi)=a_{\bar{A}}$. 

The controlled path-swap operator is then realized by:
\begin{equation}
\mathcal{U}_{\text{swap}} = \mathcal{P} \cdot \left( \ket{0}_c\bra{0} \otimes U_{\text{MZI}}(\pi) + \ket{1}_c\bra{1} \otimes U_{\text{MZI}}(0) \right),
\end{equation}
where the compensation network $\mathcal{P} = P_{A_0}(-\pi) \otimes P_{\bar{A}_1}(-\pi/2)$ is chosen to cancel the extraneous phases from $U_{\text{MZI}}$, ensuring the overall operation fulfills the defining Eq.~\eqref{eq:uswap} and results in a clean swap of the paths for the $\ket{0}_c$ component.
The inverse operation, used for recombination at the receiver, is simply $\mathcal{U}_{\text{swap}}^\dagger$.

The action of $\mathcal{U}_\text{swap}$ is as follows: if $\ket{\psi}_c = \ket{0}_c$, the probe is directed along path $A_0$; if $\ket{\psi}_c = \ket{1}_c$, it is directed along path $A_1$. The initial joint state thus becomes:
\begin{equation}
\ket{\Psi} = \frac{1}{\sqrt{2}} \left( \ket{0}_c \otimes \ket{\psi}_{A_0 B} + \ket{1}_c \otimes \ket{\psi}_{A_1 B} \right),
\end{equation}
where $\ket{\psi}_{A_0 B}$ and $\ket{\psi}_{A_1 B}$ represent the initial entangled state distributed into the two respective paths.

\subsection{Channel Implementation and Recombination}\label{channel_imple}
Each path $A_i$ interacts with a set of auxiliary modes ($D$, $E$, $F$) representing the environment and the target, via unitary interactions $V_{A_i D}$, $V_{A_i E}$, and $V_{A_i F}$. The order of these interactions is determined by the path ($A_0$ or $A_1$), implementing the desired channel superposition (ICO or PS-DE). The environmental modes are traced out afterward, yielding the noisy channel outputs.
After these interactions, the state is $\varrho_{\text{out}}^{\text{u}}$, which contains the probe in the two physical paths $A_0$, $A_1$ and the ancilla $B$ (see Eq.~\eqref{eq:varrho} in Sec.~\ref{sec:resource}).

Here, we introduce the methodology for modeling quantum channels through auxiliary modes and unitary interactions. 
We assume that the loss channel can be constructed by introducing a unitary operation $V_{A(D/F)}$ (we use the notion $D/F$ to represent ``$D$ or $F$") on the auxiliary mode $D/F$, followed by a partial trace over $D/F$. This is formally defined as:
\begin{equation}
\begin{aligned}
    \mathcal{E}^{(D/F)}(\rho_A) &:= p\rho_A +(1-p)\ket{0}_A\bra{0}\\
    &= \mathrm{Tr}_{(D/F)} \left[ V_{A(D/F)} \left( \rho_A \otimes \ket{0}_{(D/F)}\bra{0} \right) V_{A(D/F)}^\dagger \right],
\end{aligned}
\end{equation}
where $p$ is the survival probability.
The construction of the thermal-exchange channel has been detailed in the main text. It can be physically realized by coupling the transmitted mode $A$ with the thermal mode $E$ via a beam splitter, followed by tracing out the mode $E$. The mathematical representation is:
\begin{equation}
    \mathcal{E}^{(E)}(\rho_A) = \mathrm{Tr}_E \left[ V_{AE} \left( \rho_A \otimes \rho_E \right) V_{AE}^\dagger \right],
\end{equation}
where $V_{AE}$ denotes the beam-splitter unitary operation and $\rho_E$ is the thermal state of mode $E$.

We apply the aforementioned channel to the transmitted mode. Based on the path control mechanism via the control qubit, when the photon path is $A_0$, the overall unitary operation including auxiliary systems is given by:
\begin{equation}
    \mathcal{V}_0 := (V_{A_0 D} \otimes \mathbb{I}_{A_1 BEF}) (V_{A_0 E} \otimes \mathbb{I}_{A_1 BDF}) (V_{A_0 F} \otimes \mathbb{I}_{A_1 BDE}),
\end{equation}
whereas for path $A_1$, the corresponding unitary operation becomes:
\begin{equation}
    \mathcal{V}_1 := (V_{A_1 F} \otimes \mathbb{I}_{A_0 BDE}) (V_{A_1 E} \otimes \mathbb{I}_{A_0 BDF}) (V_{A_1 D} \otimes \mathbb{I}_{A_0 BEF}).
\end{equation}
Consequently, the composite unitary operation with indefinite causal order (ICO) is defined as:
\begin{equation}
    \mathcal{V}^{\text{ICO}} := \ket{0}_c\bra{0} \otimes \mathcal{V}_0 + \ket{1}_c\bra{1} \otimes \mathcal{V}_1.
\end{equation}
The output quantum state at the end of the  paths is obtained by tracing out the auxiliary systems $D, E, F$:
\begin{equation}\label{eq:ico_unitary}
    \varrho^{\text{ICO}}_{\mathrm{out}} = \mathrm{Tr}_{DEF} \left[ \mathcal{V}^{\text{ICO}} \left( \varrho_{cA_0A_1B} \otimes \ket{0}_D\bra{0} \otimes \rho_E \otimes \ket{0}_F\bra{0} \right) \mathcal{V}^{\text{ICO}\dagger} \right],
\end{equation}
where $\varrho_{cA_0A_1B}:=\ket{\Psi}_{cA_0A_1B}\bra{\Psi}$.

The final step is to recombine the paths at the receiver coherently. This is done by applying the inverse of the initial controlled path-swap operation, $\mathcal{U}_\text{swap}^\dagger$,  which maps the physical paths $(A_0, A_1)$ back to the output measurement modes $(A, \bar{A})$. The final state for measurement is obtained by tracing out the unused output mode $\bar{A}$:
\begin{equation}\label{eq:real_out}
\rho_{\text{out}}^{\text{ICO}} = \Tr_{\bar{A}}\left( \mathcal{U}_{\text{swap}}^\dagger \, \varrho_{\text{out}}^{\text{ICO}} \, \mathcal{U}_{\text{swap}} \right).
\end{equation}

\textbf{We now prove the modes $A_0$ and $A_1$ can be recombined into a single mode. Specifically, with the action of the unitary operators $\mathcal{U}^\dagger_\text{swap}$, the output state $\varrho_{\text{out}}^{\text{ICO}}$ can be transformed into the form in Eq.~\eqref{eq:real_out}.}

\begin{proof}\label{pro:a0_a1}
The proof establishes the equivalence $\mathcal{U}_\text{swap}^\dagger\varrho_\text{out}^\text{ICO}\mathcal{U}_\text{swap}=\mathcal{U}^\dagger\mathcal{P}^\dagger \varrho_{\text{out}}^{\text{ICO}} \mathcal{P}\mathcal{U} = \rho_{\text{out}}^{\text{ICO}} \otimes \ket{0}_{\bar{A}}\bra{0}$ by examining the transformation of the state's matrix elements. The key observation is the action of the combined unitary $\mathcal{U}^\dagger\mathcal{P}^\dagger(\cdot)\mathcal{PU}$ on the operators that generate the joint state of the control qubit and the two paths ($A_0$, $A_1$).

The transformation rules for the creation operators are given by:
\begin{equation}
\begin{aligned}
\mathcal{U}^\dagger\mathcal{P}^\dagger (\ket{0}_c\bra{0}\otimes a_{A_0}^\dagger)\mathcal{PU} &= \ket{0}_c\bra{0}\otimes a_A^\dagger, \\
\mathcal{U}^\dagger\mathcal{P}^\dagger (\ket{1}_c\bra{1}\otimes a_{A_1}^\dagger)\mathcal{PU} &= \ket{1}_c\bra{1}\otimes a_A^\dagger, \\
\mathcal{U}^\dagger\mathcal{P}^\dagger (\ket{0}_c\bra{0}\otimes a_{A_1}^\dagger)\mathcal{PU} &= \ket{0}_c\bra{0}\otimes a_{\bar{A}}^\dagger, \\
\mathcal{U}^\dagger\mathcal{P}^\dagger (\ket{1}_c\bra{1}\otimes a_{A_0}^\dagger)\mathcal{PU} &= \ket{1}_c\bra{1}\otimes a_{\bar{A}}^\dagger.
\end{aligned}
\end{equation}

These rules imply that the unitary $\mathcal{U}^\dagger\mathcal{P}^\dagger(\cdot)\mathcal{PU}$ effectively maps the physical path operators ($A_0$, $A_1$) to the measurement mode operators ($A$, $\bar{A}$), while simultaneously preserving the control qubit's logical state. Consequently, any matrix element of the density operator $\varrho_{\text{out}}^{\text{ICO}}$ transforms such that the information in paths $A_0$ and $A_1$ is coherently transferred to the output modes $A$ and $\bar{A}$, with the vacuum component $\ket{0}_{\bar{A}}\bra{0}$ explicitly factored out.

Given that $\varrho_{\text{out}}^{\text{ICO}}$ is constructed from linear combinations of such matrix elements, the full state must satisfy the claimed equivalence, completing the proof.
\end{proof}

This final state $\rho_{\text{out}}^{\text{u}}$, containing only the control qubit, the recombined probe mode $A$, and the ancilla $B$, is then measured jointly to decide on the target's presence.

\begin{figure}[t]
    \centering
    \includegraphics[width=0.8\linewidth]{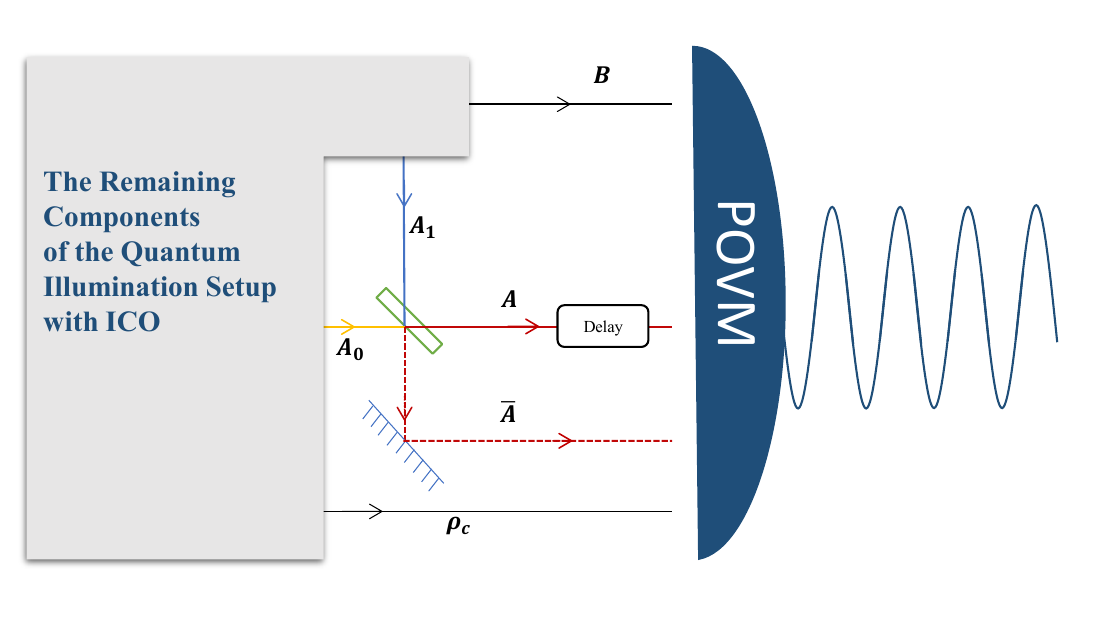}
    \caption{{\bf Robust measurement scheme for ICO quantum illumination under bit flip noise.}
    \justifying{Contrasting the original setup in Appendix~\ref{channel_imple}, where only path $A$ is measured, the proposed scheme includes simultaneous measurement of both $A$ and $\bar{A}$ to recover information that may be displaced due to bit flip errors on the control qubit.}}
    \label{fig:robustness}
\end{figure}

\subsection{Robust Measurement under Bit-Flip Noise}\label{experimental setup}
If the control qubit suffers a bit-flip error \textit{before} recombination, the probe is routed to the wrong interferometer output port ($A$ or $\bar{A}$). The standard strategy of measuring only output $A$ then results in an effective loss channel, degrading performance.

The proposed robust strategy, shown in Fig.~\ref{fig:robustness}, is to forego the final trace over $\bar{A}$ and instead perform a joint measurement on \textit{both} output ports $A$ and $\bar{A}$. This preserves the information from both paths, even if a bit-flip occurs, allowing the recovery of the quantum advantage. This adjustment aligns with the theoretical analysis in Section~\ref{sec:resource}.

\section{Derivation of the Kraus Representation for Indefinite Causal Order and Path Superposition}\label{app:Kraus}

In this appendix, we provide a detailed derivation of the Kraus operator representations for both the Indefinite Causal Order (ICO) and Path Superposition with disjoint environments (PS-DE) protocols. The key difference lies in their respective global unitary operators and the environmental modes they interact with. The global unitary of both protocols can be expressed as
\begin{equation}
    \mathcal{V}^\text{u}=\ket{0}_c\bra{0}\otimes\mathcal{V}_0^\text{u}+\ket{1}_c\bra{1}\otimes\mathcal{V}^\text{u}_1,
\end{equation}
which acts on the Hilbert space $\mathcal{H}_c\otimes\mathcal{H}_{A_0}\otimes\mathcal{H}_{A_1}\otimes\mathcal{H}_B\otimes\mathcal{H}_\text{auxiliary(u)}\otimes\mathcal{H}_E$ with $\text{auxiliary(ICO)}=\{D,F\}$ and $\text{auxiliary(PS-DE)}=\{D,F,X,Y\}$.

The global initial state that combines the system and the environment is
\begin{equation}
\varrho_\text{init}^\text{u}=\rho_{cA_0A_1B}\otimes\ket{0}_\text{auxiliary(u)}\bra{0}\otimes\rho_E,
\end{equation}
As is mentioned in Appendix~\ref{app:experimental_setup}, the output state after the receiver is obtained by:
\begin{equation}
    \begin{aligned}    \varrho_\text{out}^\text{u}&=\Tr_{\text{auxiliary(u)},E\bar{A}}\left[\mathcal{U}_\text{swap}^\dagger\mathcal{V}^\text{u}\varrho_\text{init}^\text{u}\mathcal{V}^{\text{u}\dagger}\mathcal{U}_\text{swap}\right]\\ 
    &=\Tr_{\text{auxiliary(u)},E\bar{A}}\left[\left(\mathcal{U}_\text{swap}^\dagger\mathcal{V}^\text{u}\mathcal{U}_\text{swap}\right)\left(\mathcal{U}_\text{swap}^\dagger\varrho_\text{init}^\text{u}\mathcal{U}_\text{swap}\right)\left(\mathcal{U}_\text{swap}^\dagger\mathcal{V}^{\text{u}\dagger}\mathcal{U}_\text{swap}\right)\right]\\
    &=\Tr_{\text{auxiliary(u)},E\bar{A}}\left[V^\text{u}\rho_\text{init}^\text{u}  V^{\text{u}\dagger}\otimes\ket{0}_{\bar{A}} \bra{0}\right]\\    &=\Tr_{\text{auxiliary(u)},E}\left[V^\text{u}\rho_\text{init}^\text{u}V^{\text{u}\dagger}\right],
    \end{aligned}
\end{equation}
where $\rho_\text{init}^\text{u}$ is in the Hilbert space without mode $A_0$ and $V^\text{u}$ acts on the Hilbert space without mode $A_0$.

\subsection{Kraus Representation for ICO}

Consider all auxiliary modes $D, F$, and environmental modes. The global unitary for the ICO protocol is given by:
\begin{equation}
V^{\text{ICO}} = \ket{0}_c\bra{0} \otimes V_0^\text{ICO} + \ket{1}_c\bra{1} \otimes V_1^\text{ICO},
\end{equation}
where
\begin{equation}
\begin{aligned}
V_0^\text{ICO} &:= (V_{AD} \otimes \mathbb{I}_{A B E F}) (V_{A E} \otimes \mathbb{I}_{A B D F}) (V_{A F} \otimes \mathbb{I}_{A B D E}), \\
V_1^\text{ICO} &:= (V_{A F} \otimes \mathbb{I}_{A B D E}) (V_{A E} \otimes \mathbb{I}_{A B D F}) (V_{A D} \otimes \mathbb{I}_{A B E F}).
\end{aligned}
\end{equation}

The initial state of the system plus environment is:
\begin{equation}
\rho_{\text{init}}^{\text{ICO}} = \rho_{c A B} \otimes \ket{0}_D\bra{0} \otimes \rho_E \otimes \ket{0}_F\bra{0}.
\end{equation}

The output state is obtained by:
\begin{equation}
\rho^{\text{ICO}}_{\mathrm{out}} = \mathrm{Tr}_{D E F} \left[ V^{\text{ICO}} \rho_{\text{init}}^{\text{ICO}} V^{\text{ICO}\dagger} \right].
\end{equation}

To derive the Kraus representation, we insert the completeness relation for the environmental modes:
\begin{equation}
\mathbb{I}_{D E F} = \sum_{l m n} \ket{l m n}_{D E F} \bra{l m n}.
\end{equation}

The detailed derivation proceeds as follows:
\begin{equation}
\begin{aligned}
    \rho^{\text{ICO}}_{\mathrm{out}} &= \sum_{ijk} \bra{ijk}_{DEF} V^{\text{ICO}} \left( \sum_{lmn} \ket{lmn}_{DEF}\bra{lmn} \right) \left( \rho_{cAB} \otimes \ket{0}_D\bra{0} \otimes \rho_E \otimes \ket{0}_F\bra{0} \right) \\
    &\quad \times \left( \sum_{l'm'n'} \ket{l'm'n'}_{DEF}\bra{l'm'n'} \right) V^{\text{ICO}\dagger} \ket{ijk}_{DEF} \\
    &= \sum_{ijk} \sum_{lmn} \sum_{l'm'n'} \sum_r \delta_{l0} \delta_{l'0} \delta_{mr} \delta_{m'r} \delta_{n0} \delta_{n'0} \frac{N^r}{(N+1)^{r+1}} \\
    &\quad \times \bra{ijk}_{DEF} V^{\text{ICO}} \ket{lmn}_{DEF} (\rho_{cAB}) \bra{l'm'n'}_{DEF} V^{\text{ICO}\dagger} \ket{ijk}_{DEF}\\
    &=\sum_{ijrk} \frac{N^r}{(N+1)^{r+1}} \bra{ijk}_{DEF} V^{\text{ICO}}\ket{0r0}_{DEF} (\rho_{cAB}) \bra{0r0}_{DEF} V^{\text{ICO}\dagger} \ket{ijk}_{DEF},
\end{aligned}
\end{equation}
where we have used the expansion of the thermal state $\rho_E$ that is expressed in Eq.~\eqref{eq:th}.

We define the following Kraus operators:
\begin{equation}
\begin{aligned}
    K_{i}^{(D)} &:= \mathrm{Tr}_{\tilde{A}} \left[ (\mathbb{I}_{\tilde{D}} \otimes \ket{0}_D\bra{i}) V_{AD} \right], \\
    K_{ij}^{(E)} &:= \sqrt{\frac{N^j}{(N+1)^{j+1}}} \, \mathrm{Tr}_{\tilde{A}} \left[ (\mathbb{I}_{\tilde{E}} \otimes \ket{j}_E\bra{i}) V_{AE} \right], \\
    K_{i}^{(F)} &:= \mathrm{Tr}_{\tilde{F}} \left[ (\mathbb{I}_{\tilde{F}} \otimes \ket{0}_F\bra{i}) V_{AF} \right],
\end{aligned}
\end{equation}
where $\mathbb{I}_{\tilde{X}}$ denotes the identity operator acting on all subspaces $\{D, E, F,A\}$ except the mode $X\in\{D, E, F,A\}$ and the control qubit.

It can be verified through direct computation that the quantum channels admit the following Kraus representations:
\begin{equation}
\begin{aligned}
    \mathcal{E}^{(D/F)}(\cdot) &= \sum_i K_{A,i}^{(D/F)} \, (\cdot) \, K_{A,i}^{(D/F)\dagger}, \\
    \mathcal{E}^{(E)}(\cdot) &= \sum_{jr} K_{A,jr}^{(E)} \, (\cdot) \, K_{A,jr}^{(E)\dagger}.
\end{aligned}
\end{equation}

Following the derivation above, we obtain the Kraus operators:
\begin{equation}
W_{ij r k}^{\text{ICO}} = \ket{0}_c\bra{0} \otimes \mathbb{I}_{ B} \otimes \left( K_{ i}^{(D)} K_{j r}^{(E)} K_{ k}^{(F)} \right) + \ket{1}_c\bra{1} \otimes \mathbb{I}_{B} \otimes \left( K_{ k}^{(F)} K_{ j r}^{(E)} K_{i}^{(D)} \right),
\end{equation}
 which has the form expressed in Eq.~\eqref{eq:channel_kraus}. The output state is given by:
\begin{equation}
\rho^{\text{ICO}}_{\mathrm{out}} = \sum_{i j r k} W_{ij r k}^{\text{ICO}} \rho_{c A B} W_{ij r k}^{\text{ICO}\dagger}.
\end{equation}

\subsection{Kraus Representation for PS-DE}

Consider all auxiliary modes $D, F, X, Y$, and environmental modes. The global unitary for the PS-DE protocol is fundamentally different:
\begin{equation}
V^{\text{PS-DE}} = \ket{0}_c\bra{0} \otimes V_{A Y} V_{A E} V_{A D} \otimes \mathbb{I}_{B F X} + \ket{1}_c\bra{1} \otimes V_{A X} V_{A E} V_{A F} \otimes \mathbb{I}_{B D Y}.
\end{equation}

Note that each path interacts with its own distinct set of environmental modes:
\begin{itemize}
\item Path $A_0$ ($\ket{0}_c$): interacts with modes $D$, $E$, and $Y$
\item Path $A_1$ ($\ket{1}_c$): interacts with modes $F$, $E$, and $X$
\end{itemize}

The initial state includes all environmental modes in their respective initial states:
\begin{equation}
\rho_{\text{init}}^{\text{PS-DE}} = \rho_{c A B} \otimes \ket{0}_D\bra{0} \otimes \rho_E \otimes \ket{0}_F\bra{0} \otimes \ket{0}_X\bra{0} \otimes \ket{0}_Y\bra{0}.
\end{equation}

The output state is obtained by tracing out all environmental modes:
\begin{equation}
\rho^{\text{PS-DE}}_{\mathrm{out}} = \mathrm{Tr}_{D E F X Y} \left[ V^{\text{PS-DE}} \rho_{\text{init}}^{\text{PS-DE}} V^{\text{PS-DE}\dagger} \right].
\end{equation}

We now insert completeness relations for all environmental modes:
\begin{equation}
\mathbb{I}_{D E F X Y} = \sum_{l m n o p} \ket{l m n o p}_{D E F X Y} \bra{l m n o p}.
\end{equation}

The derivation proceeds as follows:
\begin{equation}
\begin{aligned}
\rho^{\text{PS-DE}}_{\mathrm{out}} &= \sum_{i j k i' j'} \bra{i j k i' j'}_{D E F X Y} V^{\text{PS-DE}} \left( \sum_{l m n o p} \ket{l m n o p}\bra{l m n o p} \right) \rho_{\text{init}}^{\text{PS-DE}} \\
&\quad \times \left( \sum_{l' m' n' o' p'} \ket{l' m' n' o' p'}\bra{l' m' n' o' p'} \right) V^{\text{PS-DE}\dagger} \ket{i j k i' j'}_{D E F X Y} \\
&= \sum_{i j k i' j'} \sum_{l m n o p} \sum_{l' m' n' o' p'} \sum_r \delta_{l0} \delta_{l'0} \delta_{mr} \delta_{m'r} \delta_{n0} \delta_{n'0} \delta_{o0} \delta_{o'0} \delta_{p0} \delta_{p'0} \frac{N^r}{(N+1)^{r+1}} \\
&\quad \times \bra{i j k i' j'} V^{\text{PS-DE}} \ket{l m n o p} \rho_{c A B} \bra{l' m' n' o' p'} V^{\text{PS-DE}\dagger} \ket{i j k i' j'}.
\end{aligned}
\end{equation}

The numerous Kronecker delta symbols ($\delta_{l0}$, $\delta_{l'0}$, etc.) arise precisely because all auxiliary modes except $E$ are initialized in the vacuum state $\ket{0}$. These constraints enforce that only those terms where the environmental modes begin and end in the vacuum state contribute to the trace. This reflects the physical intuition that any photon loss to these environmental modes would destroy the coherence necessary for interference.

After applying these constraints, the summation simplifies dramatically:
\begin{equation}
\rho^{\text{PS-DE}}_{\mathrm{out}} = \sum_{i j r k i' j'} \frac{N^r}{(N+1)^{r+1}} \bra{i j k i' j'} \mathcal{V}^{\text{PS-DE}} \ket{0 r 0 0 0} \rho_{c A B} \bra{0 r 0 0 0} \mathcal{V}^{\text{PS-DE}\dagger} \ket{i j k i' j'}.
\end{equation}

We now define the Kraus operators for the PS-DE protocol. For the path $A_0$ (control state $\ket{0}_c$), the relevant operators are:
\begin{equation}
\begin{aligned}
K_{i}^{(D)} &:= \mathrm{Tr}_{\tilde{A}} \left[ (\mathbb{I}_{\tilde{D}} \otimes \ket{0}_D\bra{i}) V_{A D} \right], \\
K_{ j r}^{(E)} &:= \sqrt{\frac{N^r}{(N+1)^{r+1}}} \, \mathrm{Tr}_{\tilde{A}} \left[ (\mathbb{I}_{\tilde{E}} \otimes \ket{r}_E\bra{j}) V_{A E} \right], \\
K_{ k}^{(F)} &:= \mathrm{Tr}_{\tilde{F}} \left[ (\mathbb{I}_{\tilde{F}} \otimes \ket{0}_F\bra{k}) V_{A F} \right].
\end{aligned}
\end{equation}

For path $A_1$ (control state $\ket{1}_c$), the relevant operators are:
\begin{equation}
\begin{aligned}
K_{ i'}^{(X)} &:= \mathrm{Tr}_{\tilde{A}} \left[ (\mathbb{I}_{\tilde{X}} \otimes \ket{0}_X\bra{i'}) V_{A X} \right], \\
K_{j r}^{(E)} &:= \sqrt{\frac{N^r}{(N+1)^{r+1}}} \, \mathrm{Tr}_{\tilde{A}} \left[ (\mathbb{I}_{\tilde{E}} \otimes \ket{r}_E\bra{j}) V_{A E} \right], \\
K_{ j'}^{(Y)} &:= \mathrm{Tr}_{\tilde{Y}} \left[ (\mathbb{I}_{\tilde{Y}} \otimes \ket{0}_Y\bra{j'}) V_{A Y} \right].
\end{aligned}
\end{equation}

The composite Kraus operator for the PS-DE protocol is then:
\begin{equation}
\begin{aligned}
W_{i j r k i' j'}^{\text{PS-DE}} &:= \bra{i j k i' j'} V^{\text{PS-DE}} \ket{0 r 0 0 0} \\
&= \ket{0}_c\bra{0} \otimes \mathbb{I}_{ B} \otimes \left( K_{ k}^{(Y)} K_{j r}^{(E)} K_{ i}^{(D)} \right) \otimes \bra{i' j'}_{FX} \mathbb{I}_{FX } \ket{0 0}_{FX } \\
&\quad + \ket{1}_c\bra{1} \otimes \mathbb{I}_{ B} \otimes \left( K_{ j'}^{(X)} K_{ j r}^{(E)} K_{i'}^{(F)} \right) \otimes \bra{i j}_{D Y} \mathbb{I}_{D Y} \ket{0 0}_{D Y} \\
&= \ket{0}_c\bra{0} \otimes \mathbb{I}_{ B} \otimes \left( K_{ k}^{(Y)} K_{ j r}^{(E)} K_{ i}^{(D)} \right) \delta_{i'0} \delta_{j'0} \\
&\quad + \ket{1}_c\bra{1} \otimes \mathbb{I}_{ B} \otimes \left( K_{A j'}^{(X)} K_{j r}^{(E)} K_{i'}^{(F)} \right) \delta_{i0} \delta_{k0}.
\end{aligned}
\end{equation}

The Kronecker delta symbols $\delta_{i'0}$, $\delta_{j'0}$, $\delta_{i0}$, and $\delta_{k0}$ explicitly appear here due to the vacuum initial states of modes $X$, $Y$, $D$, and $F$, respectively. These constraints enforce that for the path $A_0$ to contribute, modes $F$ and $X$ must remain in their vacuum states ($i' = j' = 0$), and similarly for the path $A_1$ to contribute, modes $D$ and $Y$ must remain in their vacuum states ($i = k = 0$). Similarly,  $W_{ijkri'j'}^\text{PS-DE}$ has the form given in Eq.~\eqref{eq:channel_kraus}.
The output state for the PS-DE protocol is therefore:
\begin{equation}
\rho^{\text{PS-DE}}_{\mathrm{out}} = \sum_{i j r k i' j'} W_{i j r k i' j'}^{\text{PS-DE}} \rho_{c A B} W_{i j r k i' j'}^{\text{PS-DE}\dagger}.
\end{equation}

\section{Detailed Proof of Inequality Eq.~\eqref{eq:ineq}}\label{app:a}
For convenience, we're going to use $\rho$ for $\rho_{\mathrm{\text{out}}}^\text{QI}(\eta)$, $\rho'$ for $\rho_{\mathrm{\text{out}}}^\text{QI}(0)$, and $\sigma$ for $\sigma^\text{ICO}(\eta))$, $\sigma'$ for $\sigma^\text{ICO}(0))$. 
To compare $\epsilon^{\text{ICO}}$ and $\epsilon$, we employ key theoretical results concerning matrix trace inequalities and Lieb's concavity theorem, which are pivotal in our analysis.

\begin{proof}
\textit{Lieb's Concavity Theorem}~\cite{Lieb1973267,FAWZI2017240,Carlen2009TRACEIA} states that for any pair of positive definite matrices $A$ and $B$ and for $0\le s\le 1$, the function $f(A,B)=\tr(A^sB^{1-s})$ is jointly concave. This implies that for any $\lambda\in[0,1]$, we can use the \textit{Jensen's inequality}~\cite{Bhatia1997}
\begin{equation}\label{eq:jensen}
    f(\lambda A+(1-\lambda)C,\lambda B+(1-\lambda)D)\ge \lambda f(A,B)+(1-\lambda)f(C,D),
\end{equation}
with semi-definite positive matrices $C$ and $D$.

Let $\rho$ and $\rho'$ denote the initial density matrices of the system, with
$\sigma$ and $\sigma'$ are additional terms introduced by the ICO framework. 
Given that $\rho-\sigma$ is semi-definite positive and $0\le s\le 1$, we proceed to compare $\epsilon^{\text{ICO}}$ and $\epsilon^{QI}$.

Given the expressions for $\epsilon^{\text{ICO}}$ and $\epsilon^{\text{QI}}$ we analyze their relationship through the lens of Lieb's concavity theorem.
For $\epsilon^{\text{ICO}}$, considering the terms $(\rho+\sigma)^s(\rho'+\sigma')^{1-s}$ and $(\rho-\sigma)^s(\rho'-\sigma')^{1-s}$, each term can be viewed as an application of the function 
$f$ to different matrix pairs within the context of ICO.
Applying \textit{Lieb's Concavity Theorem} and \textit{Jensen's inequality} to each term separately:
\begin{equation}
    f(\rho,\rho')=\Tr(\rho^s\rho'^{1-s}),
\end{equation}
Since $\rho=\frac{\rho+\sigma}{2}+\frac{\rho-\sigma}{2}$ and $\rho'=\frac{\rho'+\sigma'}{2}+\frac{\rho'-\sigma'}{2}$, i.e., let $A=\rho+\sigma$, $C=\rho-\sigma$, $B=\rho'+\sigma'$, $D=\rho'-\sigma'$ and $\lambda=\frac{1}{2}$ we get
\begin{equation}
\begin{split}
    f(\rho,\rho')&=f[\frac{\rho+\sigma}{2}+\frac{\rho-\sigma}{2},\frac{\rho'+\sigma'}{2}+\frac{\rho'-\sigma'}{2}]\\
    &\ge \frac{1}{2}f(\rho+\sigma,\rho'+\sigma')+\frac{1}{2}f(\rho-\sigma,\rho'-\sigma')\\
    &=\frac{1}{2}\Tr[(\rho+\sigma)^s(\rho'+\sigma')^{1-s}+(\rho-\sigma)^s(\rho'-\sigma')^{1-s}]\\
    &=\Tr[\rho^{\text{ICO}}_\text{out}(\eta)^s\rho^{\text{ICO}}_\text{out}(0)^{1-s}].
\end{split}    
\end{equation}
Let $s^*={\arg\min}_{0\le s\le 1}\Tr[\rho_\text{out}^\text{QI}(\eta)^s\rho_\text{out}^\text{QI}(0)^{1-s}]$, we obtain:
\begin{equation}
    \Tr[\rho^{\text{ICO}}_\text{out}(\eta)^{s^*}\rho^{\text{ICO}}_\text{out}(0)^{1-{s^*}}]\le \Tr[\rho_\text{out}^\text{QI}(\eta)^{s^*}\rho_\text{out}^\text{QI}(0)^{1-{s^*}}].
\end{equation}
Thus we get 
\begin{equation}
\min_{0\le s \le 1}\Tr[\rho^{\text{ICO}}_{\text{\text{out}}}(\eta)^s\rho^{\text{ICO}}_{\text{\text{out}}}(0)^{1-s}]\leq \Tr[\rho^{\text{ICO}}_{\text{\text{out}}}(\eta)^{s^*}\rho^{\text{ICO}}_{\text{\text{out}}}(0)^{1-{s^*}}]\leq \min_{0\leq s\leq 1}\Tr \left[\rho_{\text{\text{out}}}^\text{QI}(\eta)^s\rho_{\text{\text{out}}}^{QI}(0)^{1-s}\right].
\end{equation}
Further, we can get the inequality Eq.~\eqref{eq:ineq}.
\end{proof}

\section{Proof of Theorem \ref{th:gpm}}\label{app:gpm}
\begin{proof}
We begin by defining the auxiliary functions:
\begin{equation}
\begin{aligned}
g_+(\rho,\rho',\sigma, \sigma') &:=\Tr(\rho^s\rho'^{1-s})-\Tr\left[(\rho+\sigma)^s(\rho'+\sigma')^{1-s}\right], \\
g_-(\rho,\rho',\sigma, \sigma') &:=\Tr(\rho^s\rho'^{1-s})-\Tr\left[(\rho-\sigma)^s(\rho'-\sigma')^{1-s}\right].
\end{aligned}
\end{equation}
The function of interest is then given by the average:
\begin{equation}
g_s(\rho,\rho',\sigma, \sigma')=\frac{1}{2}\left[g_+(\rho,\rho',\sigma, \sigma')+g_-(\rho,\rho',\sigma, \sigma')\right].
\end{equation}

We now bound $|g_+|$. By adding and subtracting terms, we write:
\begin{equation}
g_+(\rho,\rho',\sigma, \sigma') = \Tr\left\{\left[(\rho+\sigma)^s-\rho^s\right](\rho'+\sigma')^{1-s}+ \rho^s\left[(\rho'+\sigma')^{1-s}-\rho'^{1-s}\right]\right\}.
\end{equation}
Applying the triangle inequality and the Hölder-type inequality for the trace norm (i.e., $|\Tr(AB)| \leq \|A\|{\text{spec}} \cdot \|B\|_\text{HS}$, and noting that $\|B\|_\text{HS} \leq \mathcal{D} \|B\|{\text{spec}}$ for any operator $B$ on a $\mathcal{D}$-dimensional space), we obtain:
\begin{equation}
\begin{aligned}
|g_+(\rho,\rho',\sigma, \sigma')| \leq &\left|\Tr\left\{\left[(\rho+\sigma)^s-\rho^s\right](\rho'+\sigma')^{1-s}\right\}\right| + \left|\Tr\left\{\rho^s\left[(\rho'+\sigma')^{1-s}-\rho'^{1-s}\right]\right\}\right| \\
\leq & \mathcal{D} \left( \|(\rho+\sigma)^s-\rho^s\|_{\text{spec}} \cdot \|(\rho'+\sigma')^{1-s}\|_{\text{spec}} + \|\rho^s\|_{\text{spec}} \cdot \|(\rho'+\sigma')^{1-s}-\rho'^{1-s}\|_{\text{spec}} \right).
\end{aligned}
\end{equation}

We now bound the individual terms. For the first difference term, note that the function $f(X) = X^s$ for $s \in [0,1]$ is operator monotone and Lipschitz continuous on the cone of positive semidefinite operators (restricted to the domain of interest, e.g., operators with spectrum in $[0,1]$). Thus, there exists a Lipschitz constant $L_1(s)$ such that:
\begin{equation}
\|(\rho+\sigma)^s-\rho^s\|_{\text{spec}} \leq L_1(s) \|(\rho+\sigma) - \rho\|_{\text{spec}} = L_1(s) \|\sigma\|_{\text{spec}}.
\end{equation}
Similarly, for the function $f(X) = X^{1-s}$, there exists a Lipschitz constant $L_2(s)$ such that:
\begin{equation}
\|(\rho'+\sigma')^{1-s}-\rho'^{1-s}\|_{\text{spec}} \leq L_2(s) \|(\rho'+\sigma') - \rho'\|_{\text{spec}} = L_2(s) \|\sigma'\|_{\text{spec}}.
\end{equation}
The Lipschitz constants $L_1(s), L_2(s)$ depend on $s$ and the specific norm used; see, e.g., \cite{1981Lipschitz} for detailed bounds.

The remaining spectral norms are bounded as follows. Since $\rho'$ and $\sigma'$ are positive semidefinite, we have:
\begin{equation}
\|(\rho'+\sigma')^{1-s}\|_{\text{spec}} \leq \|\rho'+\sigma'\|_{\text{spec}}^{1-s} \leq (\|\rho'\|_{\text{spec}} + \|\sigma'\|_{\text{spec}})^{1-s} \leq (1 + \|\sigma'\|_{\text{spec}})^{1-s},
\end{equation}
where we used the subadditivity of the spectral norm and the fact that $\|\rho'\|{\text{spec}} \leq 1$ (as $\rho'$ is a density matrix). Similarly,
\begin{equation}
\|\rho^s\|_{\text{spec}} \leq \|\rho\|_{\text{spec}}^s \leq 1.
\end{equation}

Combining these bounds yields:
\begin{equation}
|g_+(\rho,\rho',\sigma,\sigma')| \leq \mathcal{D} \left( L_1(s) \|\sigma\|_{\text{spec}} (1 + \|\sigma'\|_{\text{spec}})^{1-s} + L_2(s) \|\sigma'\|_{\text{spec}} \right).
\end{equation}
By an identical argument, the same bound holds for $|g_-(\rho,\rho',\sigma,\sigma')|$:
\begin{equation}
|g_-(\rho,\rho',\sigma,\sigma')| \leq \mathcal{D} \left( L_1(s) \|\sigma\|_{\text{spec}} (1 + \|\sigma'\|_{\text{spec}})^{1-s} + L_2(s) \|\sigma'\|_{\text{spec}} \right).
\end{equation}

Therefore, by the triangle inequality and the definition of $g_s$, we conclude:
\begin{equation}
\begin{aligned}
|g_s(\rho,\rho',\sigma,\sigma')| &= \frac{1}{2} |g_+ + g_-| \\
&\leq \frac{1}{2} \left( |g_+| + |g_-| \right) \\
&\leq \mathcal{D} \left( L_1(s) \|\sigma\|_{\text{spec}} (1 + \|\sigma'\|_{\text{spec}})^{1-s} + L_2(s) \|\sigma'\|_{\text{spec}} \right),
\end{aligned}
\end{equation}
which completes the proof.
\end{proof}

\section{Proof of Theorem \ref{th:1}}\label{app:b}
\begin{proof}
Let $0 \leq \gamma_1 < \gamma_2 \leq 1$, and define $\lambda = \gamma_1 / \gamma_2 \in [0,1]$. It is easy to see that
\begin{equation}
\begin{aligned}
\rho + \gamma_1 \sigma &= \lambda (\rho + \gamma_2 \sigma) + (1-\lambda)\rho\\
\rho' + \gamma_1 \sigma' &= \lambda (\rho' + \gamma_2 \sigma) + (1-\lambda)\rho'\\
\rho - \gamma_1 \sigma &= \lambda (\rho - \gamma_2 \sigma) + (1-\lambda)\rho\\
\rho' - \gamma_1 \sigma' &= \lambda (\rho' - \gamma_2 \sigma') + (1-\lambda)\rho'.
\end{aligned}
\end{equation}
Let $A = \rho \pm \gamma_2 \sigma$, $B = \rho' \pm \gamma_2 \sigma'$, $C = \rho$, and $D = \rho'$. From Jensen's inequality Eq.~\eqref{eq:jensen}, we obtain
\begin{equation}
f(\rho \pm \gamma_1 \sigma, \rho' \pm \gamma_1 \sigma') \geq \lambda f(\rho \pm \gamma_2 \sigma, \rho' \pm \gamma_2 \sigma') + (1-\lambda)f(\rho, \rho').
\end{equation}
Combining these results, and recalling Eq.~\eqref{eq:fgamma}, we have
\begin{equation}
\begin{aligned}
F_s(\gamma_1) \geq \lambda F_s(\gamma_2) + (1-\lambda)f(\rho, \rho').
\end{aligned}
\end{equation}
Furthermore, by setting $A = \rho + \gamma_2 \sigma$, $B = \rho' + \gamma_2 \sigma'$, $C = \rho - \gamma_2 \sigma$, $D = \rho' - \gamma_2 \sigma'$, and $\lambda' = 1/2$, we can obtain
\begin{equation}
\begin{aligned}
    f(\rho, \rho') &=f\left[\lambda'(\rho+\gamma_2\sigma)+(1-\lambda')(\rho-\gamma_2\sigma),\lambda'(\rho'+\gamma_2\sigma')+(1-\lambda')(\rho'-\gamma_2\sigma') \right]\\
    &\geq \lambda'f(\rho+\gamma_2\sigma,\rho'+\gamma_2\sigma)+(1-\lambda')f(\rho-\gamma_2\sigma, \rho'-\gamma_2\sigma')\\
    &=\frac{1}{2}\left[f(\rho+\gamma_2\sigma,\rho'+\gamma_2\sigma)+f(\rho-\gamma_2\sigma, \rho'-\gamma_2\sigma')\right]\\
    &=F_s(\gamma_2).
\end{aligned}
\end{equation}
Therefore,
\begin{equation}
F_s(\gamma_1) \geq \lambda F_s(\gamma_2) + (1-\lambda)f(\rho, \rho') \geq \lambda F_s(\gamma_2) + (1-\lambda)F_s(\gamma_2)=F_s(\gamma_2).
\end{equation}
In summary, $F_s(\gamma)$ is a monotonically decreasing function, and the theorem is proven.
\end{proof}
\end{appendix}
\end{document}